\documentclass[aps,twocolumn,superscriptaddress]{revtex4}

\usepackage{amsmath,amssymb,graphicx}
\usepackage{color}

\graphicspath{{fig-ps/}}

\begin{document}

\title{Stability of single and multiple matter-wave dark solitons 
in collisionally inhomogeneous Bose-Einstein condensates}

\author{P.G.~Kevrekidis}
\email{kevrekid@math.umass.edu}
\affiliation{Department of Mathematics and Statistics,
University of Massachusetts,
Amherst, Massachusetts 01003-4515 USA}

\author{R.~Carretero-Gonz{\'a}lez}
\affiliation{Nonlinear Dynamical Systems
Group,\footnote{\texttt{URL}: http://nlds.sdsu.edu}
Computational Sciences Research Center, and
Department of Mathematics and Statistics,
San Diego State University, San Diego, California 92182-7720, USA}

\author{D.J.~Frantzeskakis}
\affiliation{Department of Physics, National and Kapodistrian University of Athens,
Panepistimiopolis, Zografos, Athens 15784, Greece}

\begin{abstract}
We examine the spectral properties of 
single and multiple matter-wave dark solitons in Bose-Einstein condensates confined in 
parabolic traps, where the scattering length 
is periodically modulated. In addition to the large-density limit
picture previously established for homogeneous nonlinearities,
we explore a perturbative analysis in the vicinity of the linear limit,
which provides good agreement with the observed spectral modes.
Between these two
analytically tractable limits, we use numerical computations to
fill in the relevant intermediate regime. We find that
the scattering length modulation can cause a variety of features
absent for homogeneous nonlinearities.
Among them, we note the potential oscillatory instability even of the single
dark soliton, the potential absence of instabilities in the immediate vicinity
of the linear limit for two dark solitons, and the existence of an
exponential instability associated with the in-phase motion of three dark
solitons.
\end{abstract}

\maketitle

\section{Introduction}

Over the past 20 years, the physics 
of atomic Bose-Einstein condensates (BECs)
has enabled the examination of numerous physical concepts~\cite{book1,book2} (see also 
Ref.~\cite{dumitr1} for a recent review on the progress in this setting). One of the
key directions that have been explored lies
at the interface between nonlinear wave dynamics and such
atomic (as well as optical) systems, concerning
the study of so-called matter-wave solitons~\cite{emergent,book_new}.
These coherent structures have been not only theoretically
predicted but also in numerous cases experimentally verified.
Some of the most notable relevant examples are
bright~\cite{expb1,expb2,expb3},
dark~\cite{djf} and gap~\cite{gap} matter-wave solitons.
Higher dimensional analogues of these structures have also been
studied including vortices~\cite{fetter1,fetter2},
solitonic vortices and vortex rings~\cite{komineas_rev}.

In the one-dimensional (1D), single-component repulsive BEC setting,
arguably, the most prototypical nonlinear excitations are 
dark solitons.
These states have received considerable attention,
perhaps in large measure due to their accessibility
in a diverse array of experiments featuring a wide variety
of methods. These have led to their (progressively more) well-controlled
creation~\cite{chap01:denschl,chap01:dark1,ourmarkus2,ourmarkus3,seng1,seng2}.
Additionally, numerous works have examined dark solitons in
higher-dimensional settings observing experimentally their instability in the latter settings.
This, in turn, leads to the formation of vortices, vortex rings and vortex lines, as illustrated, 
e.g., in Refs.~\cite{chap01:dark,engels,jeff,rcg:90,pitasv}.
One of the important associated recent developments has been the 
realization of such nonlinear excitations in fermionic 
superfluids~\cite{martinz,fprl}.
A relatively comprehensive summary of the relevant activity, encompassing
both experiment and theory ---although not including some of the most
recent developments--- can be found in Ref.~\cite{djf} and also in
Chap.~2 of Ref.~\cite{book_new}.

A context that has recently received considerable interest
as regards the dynamics of solitary waves is that of 
spatially inhomogeneous nonlinearities. A review exploring diverse aspects
of this topic can be found in Ref.~\cite{borisreview}. 
In BECs, the nonlinearity stems from interatomic interactions; 
in the mean-field approach, where BECs are described by a macroscopic BEC wavefunction 
obeying the Gross-Pitaevskii (GP) equation, 
the nonlinearity coefficient is proportional to the $s$-wave scattering 
length~\cite{book1,book2}. 
The fact that the sign and magnitude of 
the interatomic interactions can be controlled using
Feshbach resonances~\cite{Koehler,feshbachNa,ofr}, 
has led to a wide array of theoretical \cite{theor} and experimental possibilities, 
including the realization of matter-wave bright solitons \cite{expb1,expb2,expb3} 
or the revelation of the BEC-BCS crossover \cite{exp}. 
For instance, Feshbach resonances have been experimentally used to 
induce spatial inhomogeneities in the scattering length of Yb BECs~\cite{Tak}. 
Such {\it collisional inhomogeneities}, lead to an effective nonlinear potential, 
in addition to the customary external 
(e.g.,~parabolic) potential.
As a result, this may lead to features absent in spatially uniform
condensates~\cite{NLPots,Chang,borisreview}. Some such examples
include, but are not limited to,
adiabatic compression of matter waves~\cite{our1}, enhancement of the
transmission of matter waves through barriers~\cite{our2}, dynamical trapping
of solitary waves~\cite{our2}, delocalization transitions of matter
waves~\cite{LocDeloc}, emission of atomic 
solitons \cite{in4,fot_scat}, generation of solitons \cite{21} 
and vortex rings \cite{22}, control of Faraday waves \cite{alex}, 
and others. These types of nonlinear potentials have also led to interesting 
insights in the context of photonic structures~\cite{kominis}.

Our aim in the present work is to explore states involving one or more
dark solitons in both collisionally homogeneous and collisionally
inhomogeneous BECs. In the former setting, such states have already
been explored, e.g., in Ref.~\cite{ourmarkus3} and their spectral
analysis in the limit where the solitary waves
can be considered as particles was theoretically studied in Ref.~\cite{coles}.
However, even in that setting we offer a twist by examining the
opposite (near-linear) limit by virtue of the perturbative analysis
developed in Ref.~\cite{mprizola}. The two limits, when combined, constitute -- 
in our view -- the full set of parameter ranges that can be
analytically studied,
and the intermediate parameter range between the two is supplemented 
by means of numerical computations.
We also extend both our analytical methodology and our numerical
investigations to the case of collisionally inhomogeneous
condensates. There we find a rich variety of features that are
distinct from the homogeneous case. For instance,
the single dark soliton, which is generically stable in 1D BECs, 
can become subject to a Hopf bifurcation and an oscillatory
instability in the inhomogeneous setting.
The two-soliton state, on the other hand, is immediately subject to this
type of instability beyond the linear limit for constant
nonlinearity. The associated degeneracy can be broken for inhomogeneous
nonlinearities, and this instability may be absent in the immediate
vicinity of the limit. Finally, for three dark solitons,
an instability that did not exist in the homogeneous limit arises 
featuring exponential in-phase divergence of the three solitons 
from their equilibrium solution. 
In this case too, the theoretical analysis is supplemented by numerical
bifurcation results. We also use direct numerical simulations in order
to showcase some of the above instabilities.

Our presentation is structured as follows. First, in Sec.~\ref{setup}, we introduce the model 
and present our analytical approach.
Next,  
in the first part of Sec.~\ref{sec:results}, we provide the numerical results 
for the homogeneous case, while in the second part, we consider the inhomogeneous case. 
Finally, in Sec.~\ref{sec:conclu}, we summarize our findings and present a number of 
directions for future work.

\section{Model and Computational Setup}
\label{setup}


Our model of choice will be the quasi-1D 
Gross-Pitaevskii (GP) equation of the form~\cite{book1,book2,book_new}:
\begin{eqnarray}
i U_t = -\frac{1}{2} U_{xx} + V(x) U + g(x) |U|^2 U,
\label{eqn1}
\end{eqnarray}
where $U(x,t)$ is the macroscopic BEC wavefunction, $V(x)$ is the external
potential, and $g(x)$ is the, potentially, spatially-dependent
nonlinearity coefficient.
While improved models have been devised to 
more adequately capture the effect of BEC dimensionality -- and, particularly, 
the interplay between longitudinal and transverse directions of the BEC (see,
e.g., Refs.~\cite{salasnich,delgado}) -- the methods we propose
here would be equally applicable to the latter variants. Nevertheless, 
for simplicity and clarity of the exposition (and of
the analytical results provided), as well
as to connect to earlier literature on the subject, we opt
to present the results in the simpler GP setting.

The linear external potential will be assumed to be parabolic, in
the functional form 
\begin{equation}
V(x)=\frac{1}{2} \Omega^2 x^2,
\end{equation}
with $\Omega$ being the normalized trap strength, 
while the collisional inhomogeneity, when present, will be assumed to 
have a typical periodic modulation around a constant value (see, e.g., Ref.~\cite{borisreview}), namely:
%
\begin{equation}
g(x)=1 + g_0 \sin^2(k x).
\label{g(x)}
\end{equation}
In principle,
for $g_0$ sufficiently negative, $g$ may even change 
sign, yet here we will restrict our considerations to cases
where this does not occur. The linear limit where the
wavefunction $U \rightarrow 0$ amounts to the quantum
harmonic oscillator with eigenfunctions 
\begin{equation}
U \propto H_n(\sqrt{\Omega} x)
\exp(-\Omega x^2/2) \exp(-i E_n t),
\end{equation}
and corresponding eigenenergies 
$E_n = (n + 1/2) \Omega$, where $H_n$ is the $n$-th Hermite polynomial.

We now focus on stationary states of the nonlinear problem in the
form: $U(x,t)=e^{-i \mu t} u(x)$. It is well known that such
states bifurcate from the corresponding linear states~\cite{kat,zezy}.
Following the methodology of Ref.~\cite{mprizola}, we can use
a series expansion of the solution in the vicinity of this
linear limit, in the form:
\begin{equation}
u= \sqrt{\epsilon} u_0 +
\epsilon^{3/2} u_1 + \dots, \quad
\mu=\mu_0 + \epsilon \mu_1 + \dots.
\end{equation}
where $(\mu_0,u_0)$ correspond, respectively, to the
eigenvalue and eigenfunction of a state
at the linear limit. As a result, we find from Eq.~(\ref{eqn1}), 
at O$(\epsilon)$, the solvability condition:
\begin{eqnarray}
\mu_1= \int |u_0|^4 dx. 
\label{eqn2}
\end{eqnarray}
From this formula, we can specify $\epsilon= (\mu - \mu_0)/\mu_1$
(to leading order).

The next step is to consider 
the spectral stability i.e., the
so-called Bogolyubov-de Gennes (BdG) analysis associated with
the linearization around a stationary state. We use
the ansatz
\begin{eqnarray}
U(x,t)=e^{-i \mu t} \left[u(x) + a(x) e^{\lambda t}
+ b^{\star}(x) e^{\lambda^{\star} t} \right],
\label{decomp}
\end{eqnarray}
where $\lambda$ denotes the relevant eigenvalue, and $(a,b)^T$ is its
eigenfunction 
(star denotes complex conjugate).
This leads to the eigenfrequency matrix:
%
%
\begin{equation}
M=
\begin{pmatrix}
M_{11} & M_{12} \\
M_{21} & M_{22}
\end{pmatrix},
\label{M}
\end{equation}
with
$$
\begin{array}{rcl}
M_{11}&=&\left(-\frac{1}{2} \frac{d^2}{dx^2}  +V  +2 g(x) |u|^2-\mu\right),\\[2.0ex]
M_{12}&=& g(x) u^2,\\[1.0ex]
M_{21}&=& - g(x) u^{\star 2},\\[1.0ex]
M_{22}&=& -\left(-\frac{1}{2} \frac{d^2}{dx^2}    +V  + 2 g(x) |u|^2-\mu\right).
\end{array}
$$
Here, the eigenfrequency
$\omega$ is connected to the eigenvalue $\lambda$ through
$\lambda=i \omega$.

Now, using the expansion in powers of $\epsilon$
within the stability matrix, we obtain
\begin{equation}
M v= ({\cal H}_0 + \epsilon {\cal H}_1) v= \omega v,
\end{equation}
with
\begin{eqnarray}
v &=& \left(\begin{array}{c}
a\\
b\\
\end{array} \right) , \label{Eq:uv} \\
{\cal H}_0 &=&\left( \begin{array}{cc}
{\cal L} - \mu_0 & 0 \\
0  & \mu_0 -{\cal L} \\
    \end{array} \right),
\end{eqnarray}
where ${\cal L}=-\frac{1}{2} \frac{d^2}{dx^2}  +V(r) $, while
\begin{eqnarray}
\label{Eq:uv2}
{\cal H}_1= \left( \begin{array}{cc}
2 g(x) |u_0|^2 - \mu_1 & g(x) u_0^2 \\[1.0ex]
- g(x) (u_0^2)^{\star}  & \mu_1 - 2 g(x) |u_0|^2
    \end{array} \right).
\end{eqnarray}
A nonzero
imaginary part of $\omega$ (or, equivalently, 
a nonzero real part of $\lambda$)
in this Hamiltonian system signals the presence of
a dynamical instability.

We should note in passing that for the existence problem the
leading-order correction $u_1$ satisfies
\begin{eqnarray}
&&\mu_0 u_1 + \mu_1 u_0 = {\cal L} u_1 + g(x) u_0^3
\nonumber\\[2.0ex]
&\Rightarrow& \left({\cal L}- \mu_0\right) u_1 = \mu_1 u_0 - g(x) u_0^3 \equiv F(x).
\label{eqn5}
\end{eqnarray}
Decomposing $u_1$ into modes of the quantum harmonic oscillator
$u_1= \sum_{n \neq m} a_n v_n$ with frequency (energy) $\omega_n$,
and considering the state of interest $u_0$ to be a multi-dark
soliton 
state composed by 
$m$ dark solitons, we obtain
\begin{eqnarray}
u_1  = \sum_{n \neq m} \frac{\langle v_n, F(x)\rangle}{\omega_n - \mu_0} v_n,
\label{eqn6}
\end{eqnarray}
where $\langle f , g \rangle = \int_{-\infty}^{\infty}{f(x)g(x)\,dx}$.
%
%
While the denominator acquires a particularly simple form
given that $\omega_n= (n+1/2) \Omega$ and $\mu_0= (m+1/2) \Omega$,
hence their difference is $(n-m) \Omega$, we will not pursue
the existence problem of Eq.~(\ref{eqn6}) at higher order
further. We will instead focus on the stability problem of
Eqs.~(\ref{Eq:uv})--(\ref{Eq:uv2}) where this difference
also appears, characterizing the eigenvalues $(\pm (n-m) \Omega)$
of the stability
problem associated with the operator ${\cal H}_0$.
In the case where $n>m$, these are referred to
as positive energy eigenvalues, while if $n<m$,
they are referred to as negative energy
(or anomalous) eigenvalues/eigenmodes.

To identify the dependence of the eigenvalues on the chemical potential
parameter $\mu$ that we consider in our continuations,
one uses the degenerate perturbation theory of Ref.~\cite{mprizola},
constructing the
matrix ${\cal M}$ with elements
\begin{eqnarray}
{\cal M}_{ij}= \langle W_i | {\cal H}_1 | W_j \rangle,
\label{degen}
\end{eqnarray}
for all the pairs of $i,j$ (and eigenmodes $W_i, W_j$)
which correspond to degenerate
eigenvalues in the linear limit. The eigenfrequencies of the
matrix ${\cal M}$ will yield the corrections $\omega^{{\rm cor}}$ to the
linear limit  of the stability problem according to:
\begin{eqnarray}
\omega= (n-m) \Omega + \epsilon \omega^{{\rm cor}}.
\label{degen2}
\end{eqnarray}
Recall that here $n$ indexes the $n$-th eigenmode,
while $m$ is the index of the state of interest, i.e., $m=1$
for a single dark soliton, $m=2$ for a double dark soliton, $m=3$
for a triple dark soliton, and so on. 

Having laid out this theoretical formulation, we are now
ready to obtain specific results for the two cases at hand,
namely the collisionally homogeneous case of $g(x)=1$
and the collisionally inhomogeneous, sinusoidally varying $g(x)$.

\section{Numerical Results and Comparison to Theory
\label{sec:results}}

\subsection{Collisionally Homogeneous Case}

We start with 
the case of the single dark soliton, which it is well known to be 
spectrally stable at the level of the
GP equation~\cite{book_new,djf}.
Since $m=1$, there will be two modes of the linearization
that will be degenerate at $\omega=\Omega$, a positive
energy one, with $n=2$, and a negative energy one, with $n=0$.
However, one of these modes (the $n=2$ one) is associated with the so-called
dipolar or Kohn oscillation~\cite{book1,book2}
and hence remains invariant under variations of $\mu$. The other one (the anomalous
mode) becomes associated with the oscillation frequency of
the dark soliton in the parabolic trap, predicted in the
asymptotic limit of large chemical potential to be
$\omega=\Omega/\sqrt{2}$ (this prediction was first reported 
in Ref.~\cite{busch} and was subsequently confirmed by numerous additional
works -- cf. discussion in Refs.~\cite{book_new,djf} and references therein). 
Our theoretical analysis near the linear limit
leads to $\mu_1=3 \Omega^{1/2}/(4 \sqrt{2 \pi})$ (and from
this, we express $\epsilon=(\mu - 3 \Omega/2)/\mu_1$ for this branch).
Then, using the degenerate perturbation theory, we obtain that
the frequency associated with the anomalous mode near the linear limit is:
\begin{eqnarray}
\omega= \Omega - \frac{1}{6} \left(\mu- \frac{3}{2} \Omega \right).
\label{pred1}
\end{eqnarray}
Importantly, all higher-order modes in this case are non-degenerate 
and, hence, the corresponding eigenfrequency corrections can be computed
as scalars with $i=j=1$ in Eq.~(\ref{degen}). Following this path,
we obtain
\begin{eqnarray}
\omega= 2 \Omega - \frac{1}{12}  \left(\mu- \frac{3}{2} \Omega \right),
\label{pred2}
\end{eqnarray}
for the case of the mode with $n=2$, while finally for the mode
with $n=3$, we have:
\begin{eqnarray}
\omega= 3 \Omega - \frac{7}{32}  \left(\mu- \frac{3}{2} \Omega \right).
\label{pred3}
\end{eqnarray}

\begin{figure} 
\begin{center}
\includegraphics[width=8cm]{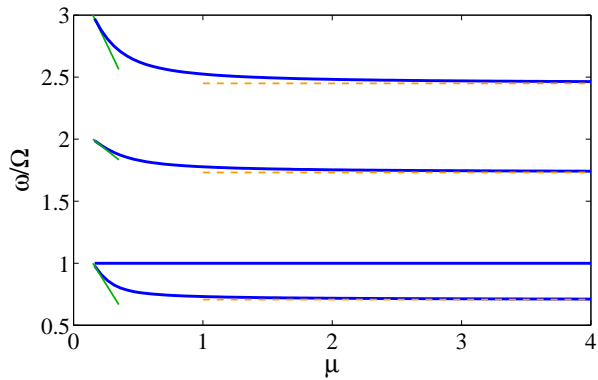}
\caption{(Color online) The modes arising from the BdG linearization
analysis around a single dark soliton. The modes are normalized
to the trap strength and the eigenfrequency
continuation is shown over the chemical
potential $\mu$, for the case of $\Omega=0.1$.
The short solid (green) lines
denote the analytical predictions near the linear limit of
Eqs.~(\ref{pred1})--(\ref{pred3}), the horizontal (orange) dashed lines
show the asymptotic limit predictions for large
$\mu$ [cf.~Eq.~(\ref{tflimit}) and $\Omega/\sqrt{2}$].
Finally, the (blue) solid lines, numerically obtained from the
full BdG problem, interpolate between these analytically 
tractable limits, providing the full spectral picture,
as it emerges through the detailed numerical BdG analysis results.}
\label{state}
\end{center}
\end{figure}

\begin{figure*}[tb] 
\begin{center}
\includegraphics[width=8cm]{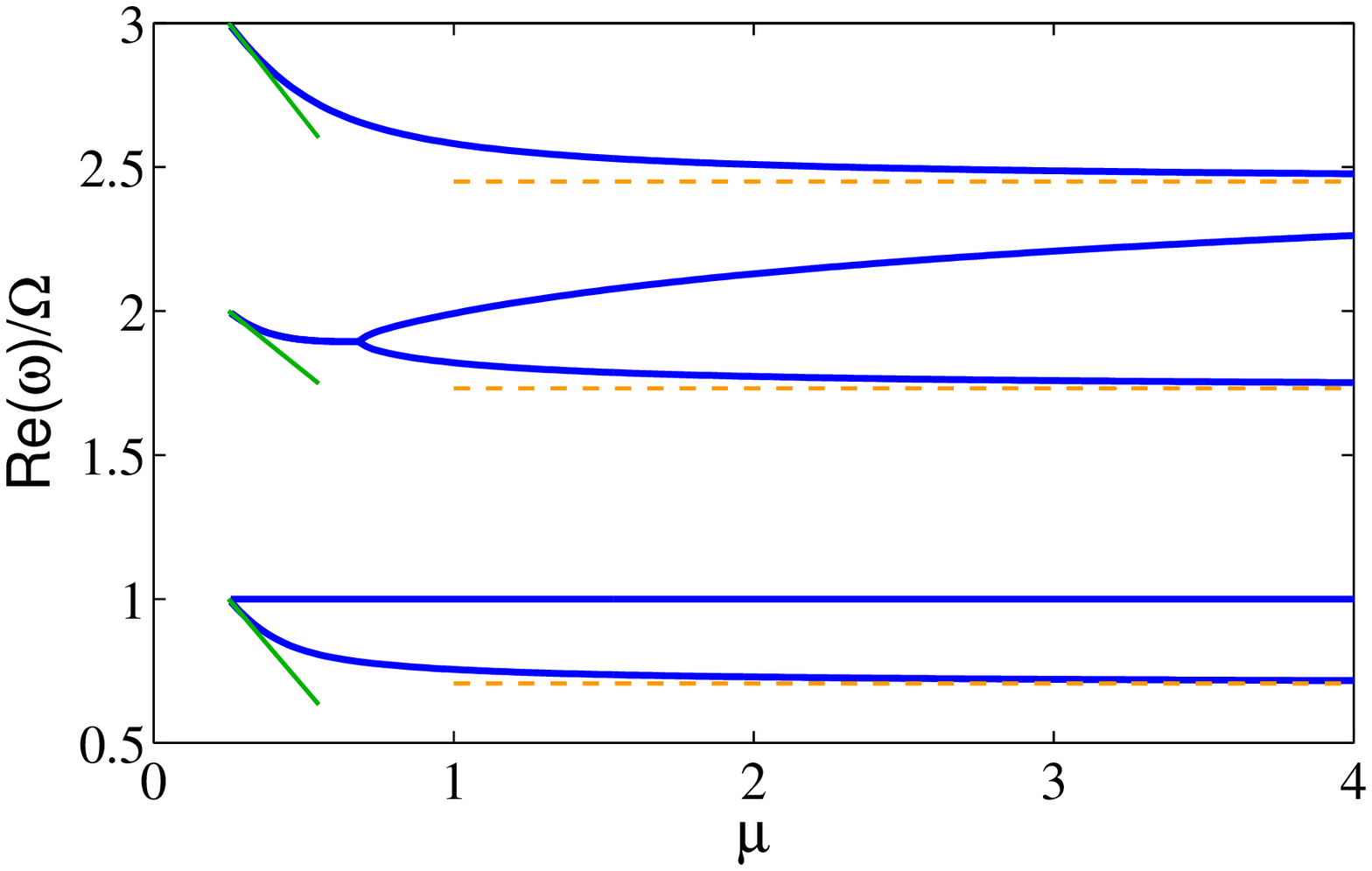}
\includegraphics[width=8cm]{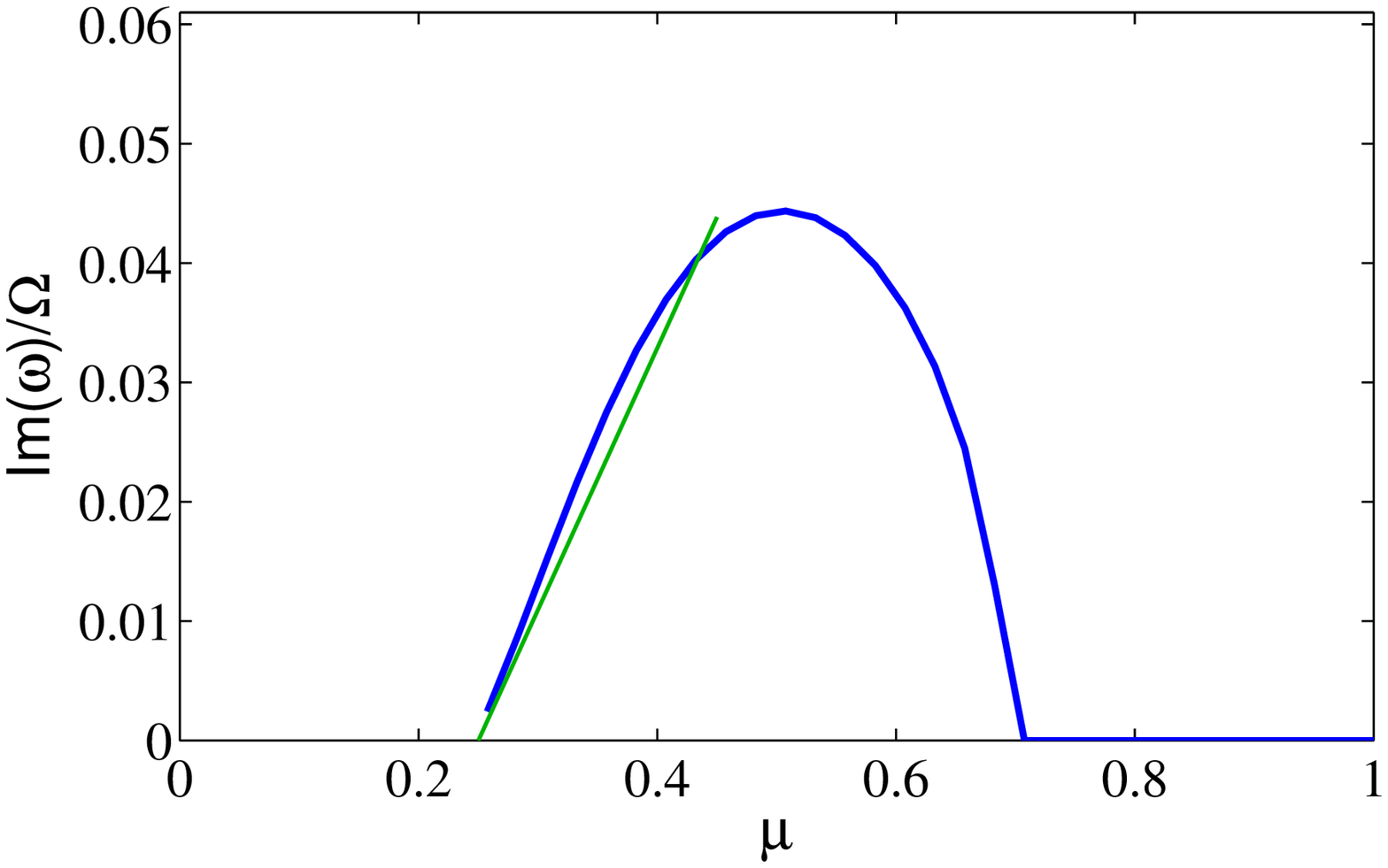} 
\caption{(Color online) The left panel is similar to 
Fig.~\ref{state},
but with the twist that we are now considering the second excited state
and the relevant theoretical predictions near the linear limit are
given by Eqs.~(\ref{pred4})--(\ref{pred6}). The right panel showcases
the imaginary part of the relevant eigenfrequencies, i.e., the growth
rate of the instability associated with the resonant modes at
$\omega=2 \Omega$. The solid (green) line represents the theoretical
prediction for this imaginary part
[cf. Eq.~(\ref{pred5})].
}
\label{LC}
\end{center}
\end{figure*}

\begin{figure*}[t] 
\begin{center}
\includegraphics[width=8cm]{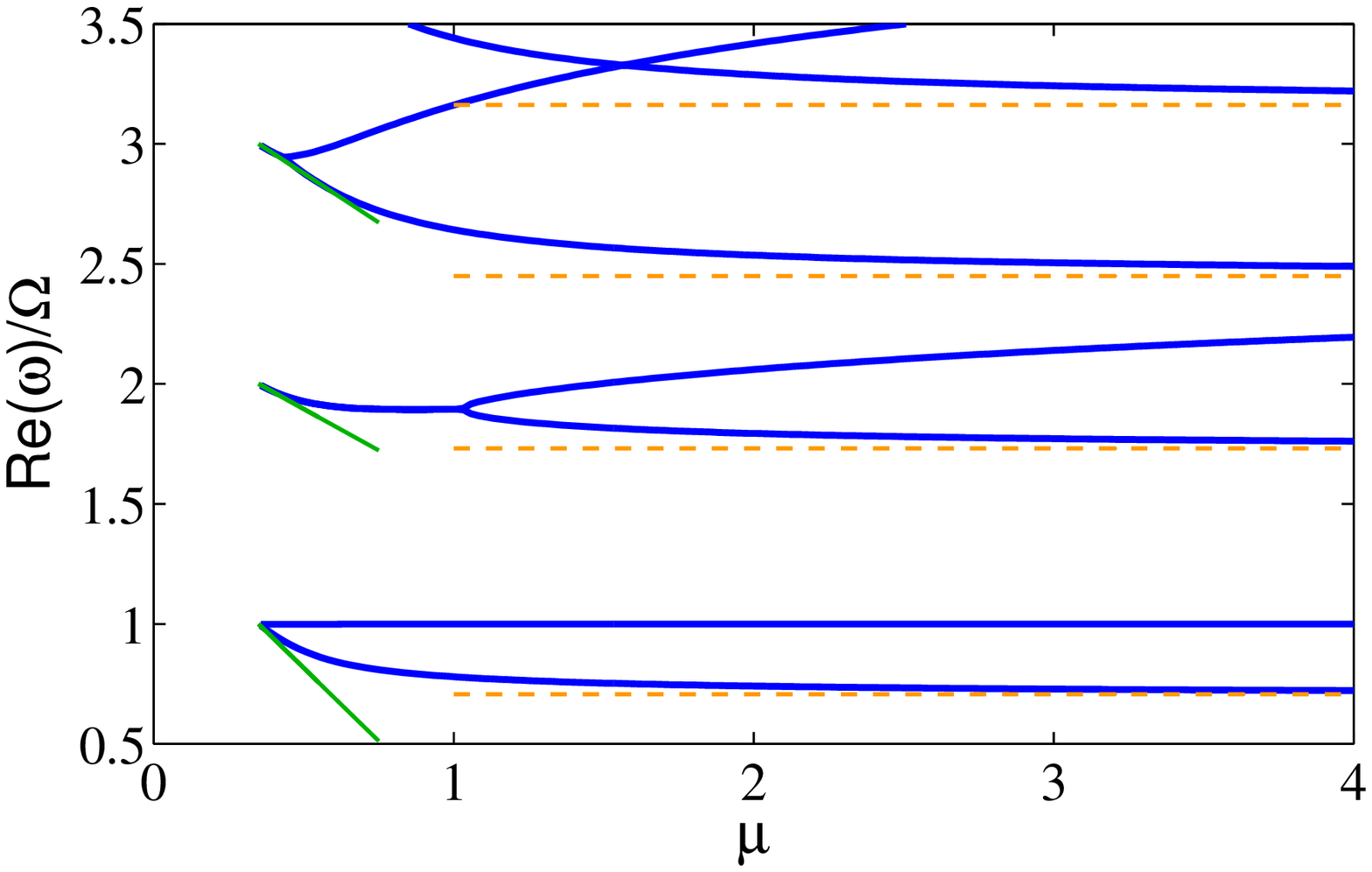}
\includegraphics[width=8cm]{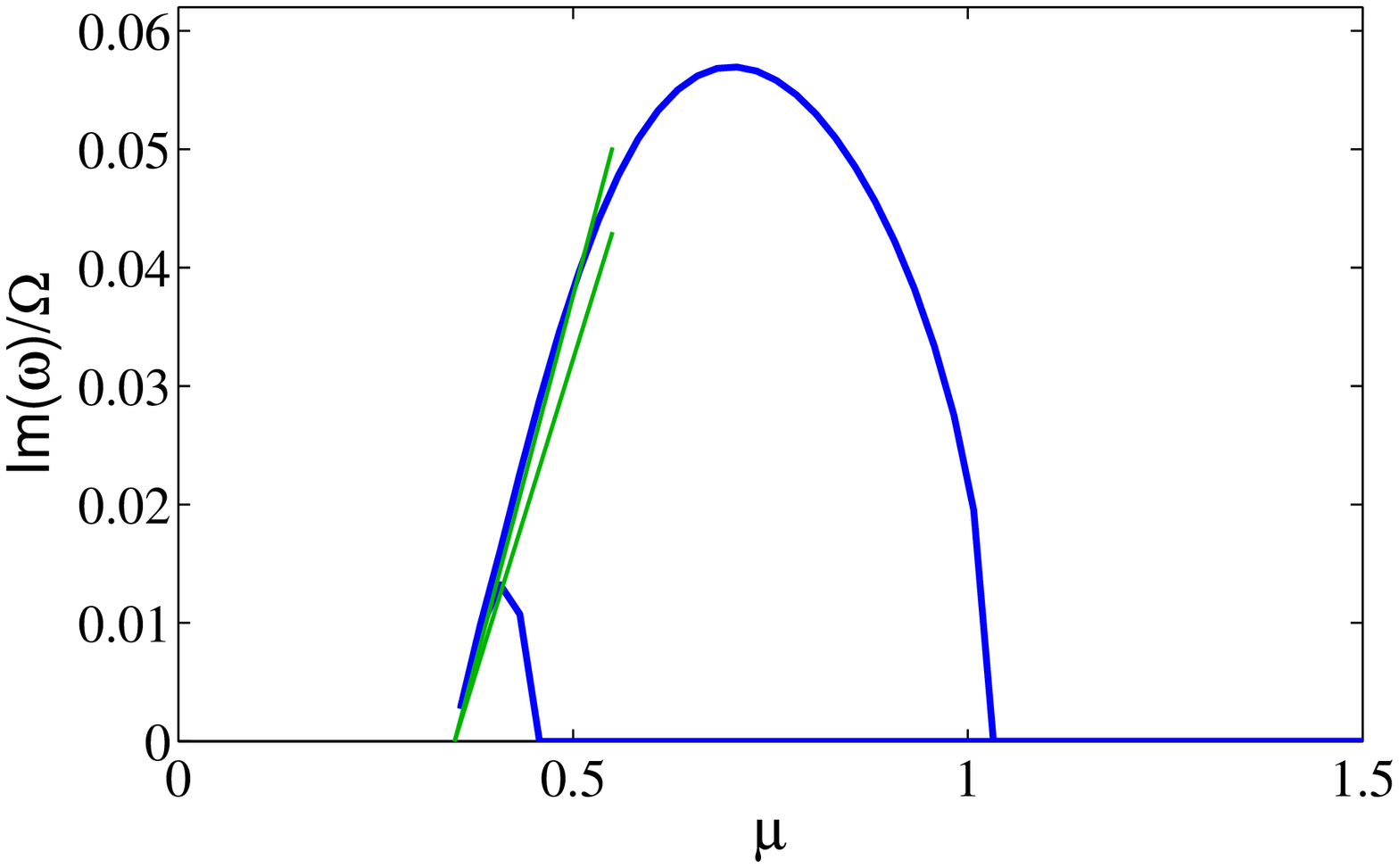}  
\caption{(Color online) 
The eigenfrequencies of the BdG analysis,
using the same notation as in Fig.~\ref{LC}, but for the 
third excited state. The main difference
from Fig.~\ref{LC} is that there are now two resonances, one occurring
at $2 \Omega$ and one at $3 \Omega$, with the corresponding imaginary
parts of the resulting instabilities shown in the right panel [from
the numerics, as well as from the imaginary part of the 
analytical predictions of Eqs.~(\ref{pred8}) and~(\ref{pred9})].
}
\label{f:spectrum}
\end{center}
\end{figure*}

The large chemical potential case also enables analytical consideration.
In this setting, the dark 
solitons become narrow
as their characteristic scale, the healing length, gets
smaller, and can be assumed as having no internal dynamics
over that scale. However, they still possess  their
own internal modes (the anomalous
or negative energy modes) that pertain to the in-trap soliton motion. 
The corresponding eigenvectors encompass evolution of the
coherent structures, but not that of the background.
In this large chemical potential, so-called Thomas-Fermi (TF) limit,
the spectrum consists of two separate ingredients. The above mentioned  
internal modes of the solitons 
and those of the background, i.e., the
ground state on top of which the solitary waves are ``supported''.
The latter modes have been characterized since
the early work of Stringari~\cite{sandro2} (see Ref.~\cite{pelirecent} for 
a recent discussion in different dimensionalities) as pertaining, through 
a suitable transformation, to Legendre differential equation and being 
eventually given by
\begin{eqnarray}
\omega_n = \sqrt{\frac{n (n+1)}{2}} \Omega.
\label{tflimit}
\end{eqnarray}
Hence, for a single dark soliton all the TF limit modes consist of
those of Eq.~(\ref{tflimit}), for non-negative
integers $n$, and the $\Omega/\sqrt{2}$ mode pertaining
to the single soliton in-trap oscillation.
The result of Eq.~(\ref{tflimit}) is obtained directly by using
the (approximate) TF solution profile $u_0= [\max(\mu-V(x),0)]^{1/2}$
in the linearization equations that get considerably simplified in this
case and, in the limit, approach the Legendre differential equation.

Both sets of predictions, namely those of Eqs.~(\ref{pred1})--(\ref{pred3}),
as well as those of Eq.~(\ref{tflimit}) (and the soliton internal mode)
are shown in Fig.~\ref{state}. The former modes can be seen to provide
a good description for small chemical potentials near the limit of
$3 \Omega/2$,
while the latter provide the
proper asymptotic limit for large $\mu$.
The numerical computations
interpolate between these two asymptotic, analytically tractable limits,
providing the full spectral picture of the BdG analysis.
It is interesting to mention here  that in the limit of
large $\mu$, we systematically observe that the convergence of the
numerical BdG results (solid blue line) to
the analytical ones of Eq.~(\ref{tflimit})
(dashed orange line)
occurs ``sooner'', i.e., for lower values of $\mu$,
for lower modes. This trait will also be discernible
in the multiple soliton cases that follow.

In the case of two dark solitons, the picture is fairly similar, however
bearing the following differences. Since now we are dealing with the $m=2$
mode, bifurcating out of $\mu_0=5 \Omega/2$, it is the modes with $n=3$
(positive energy) and $n=1$ (negative energy) that will be resonant
at $\omega=\Omega$. Again, here the $n=3$ mode corresponds to the dipolar
oscillation leading to an invariant eigenfrequency in the stability analysis.
However, the $n=1$ mode represents the lowest vibrational mode that
can be tracked via our degenerate perturbation theory, which in this
case predicts:
\begin{eqnarray}
\omega=\Omega - \frac{5}{41} \left(\mu - \frac{5}{2} \Omega \right).
\label{pred4}
\end{eqnarray}
A similar degeneracy, for $m=2$, can be diagnosed 
for the frequency 
$2 \Omega$. This is due to the fact that 
the modes with $n=4$ (positive
energy) and $n=0$ (negative energy) are degenerate. In this case,
the degenerate perturbation theory is necessary for these two modes,
resulting in a {\it complex} eigenfrequency of the form:
\begin{eqnarray}
\omega= 2 \Omega + \left(\mu - \frac{5}{2} \Omega \right)
\frac{-55 \pm 3 \sqrt{23} i }{656}.
\label{pred5}
\end{eqnarray}
Finally, in the case of the mode starting from frequency $3 \Omega$,
there is no degeneracy and the relevant mode with $n=5$
can be found to have the frequency:
\begin{eqnarray}
\omega=3 \Omega - \frac{87}{656} \left(\mu - \frac{5}{2} \Omega \right).
\label{pred6}
\end{eqnarray}

In the large chemical potential limit, the two-soliton case
naturally still bears all the modes associated with the ground
state, as per Eq.~(\ref{tflimit}). However, now there are two solitonic
modes, the lowest one still given by $\Omega/\sqrt{2}$ and associated
with the in-phase vibration of the two dark solitons (tantamount to the
oscillation of a single one inside the trap, hence bearing the same
frequency). The higher one among the two modes is associated with the
out-of-phase motion of the two solitons,  
and is provided by
Eq.~(21) in Ref.~\cite{coles}, illustrating both the single and the double
logarithmic dependence on the chemical potential $\mu$.

The numerical computations associated with this two-soliton state
are illustrated in Fig.~\ref{LC}. While the large density limit has
been explored before (e.g., in Ref.~\cite{coles}), and its accuracy is
perhaps expected, it is relevant to highlight the success of
the degenerate perturbation theory near the linear limit. The latter
captures the decreasing dependence of all of the first few modes as the
chemical potential increases. Moreover, it accurately tracks down
not only the motion of the first and third mode towards decreasing
frequencies, but also the instability caused by the degenerate second
mode at $2 \Omega$. In fact, it provides a very good quantitative handle
of the growth rate of the associated instability, as shown in the right panel
of Fig.~\ref{LC}. Once again, the numerical results interpolate between the 
analytically tractable limits in this two-soliton case.

\begin{figure*}[tb] 
\begin{center}
\includegraphics[width=8.00cm]{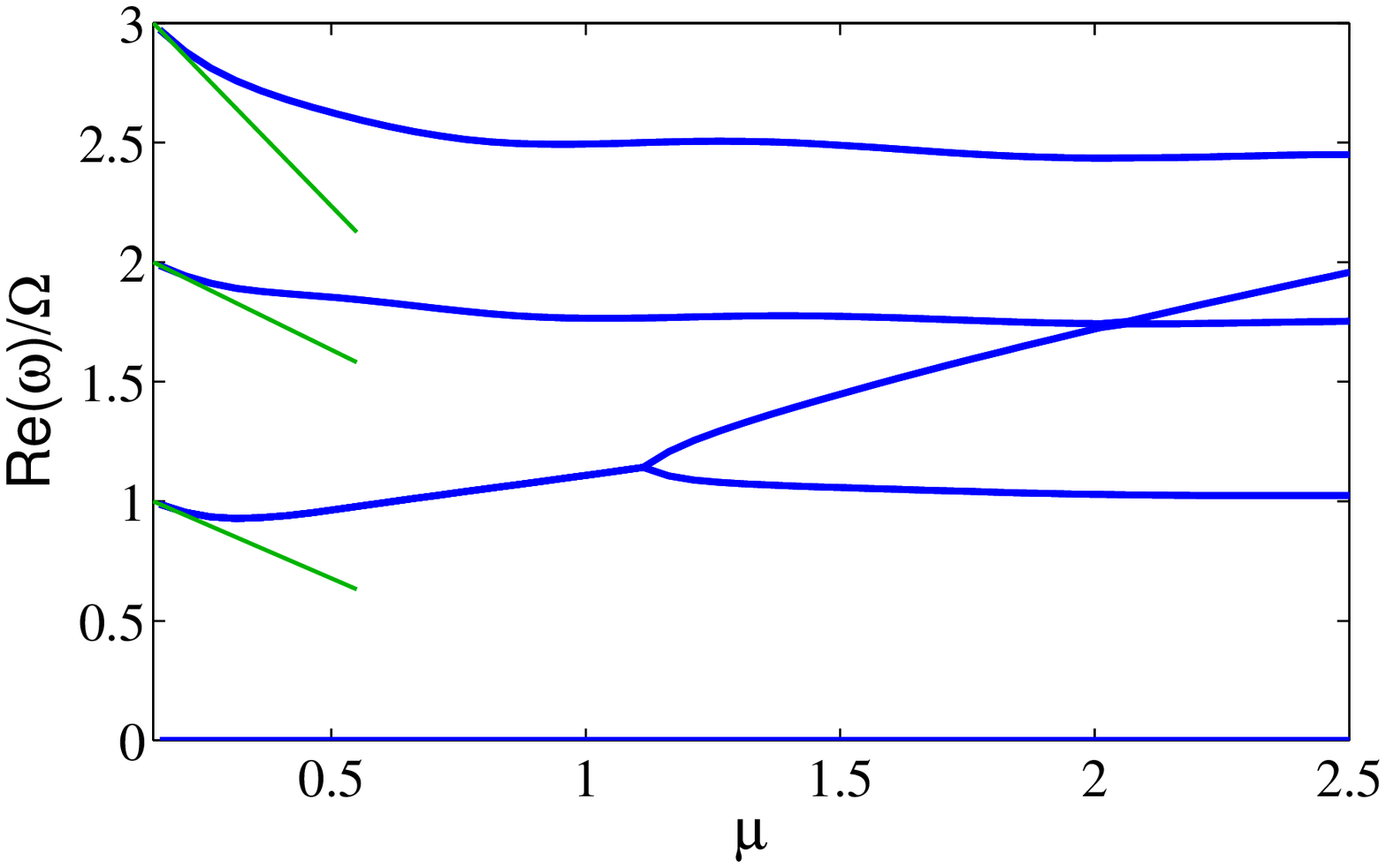}
\includegraphics[width=8.00cm]{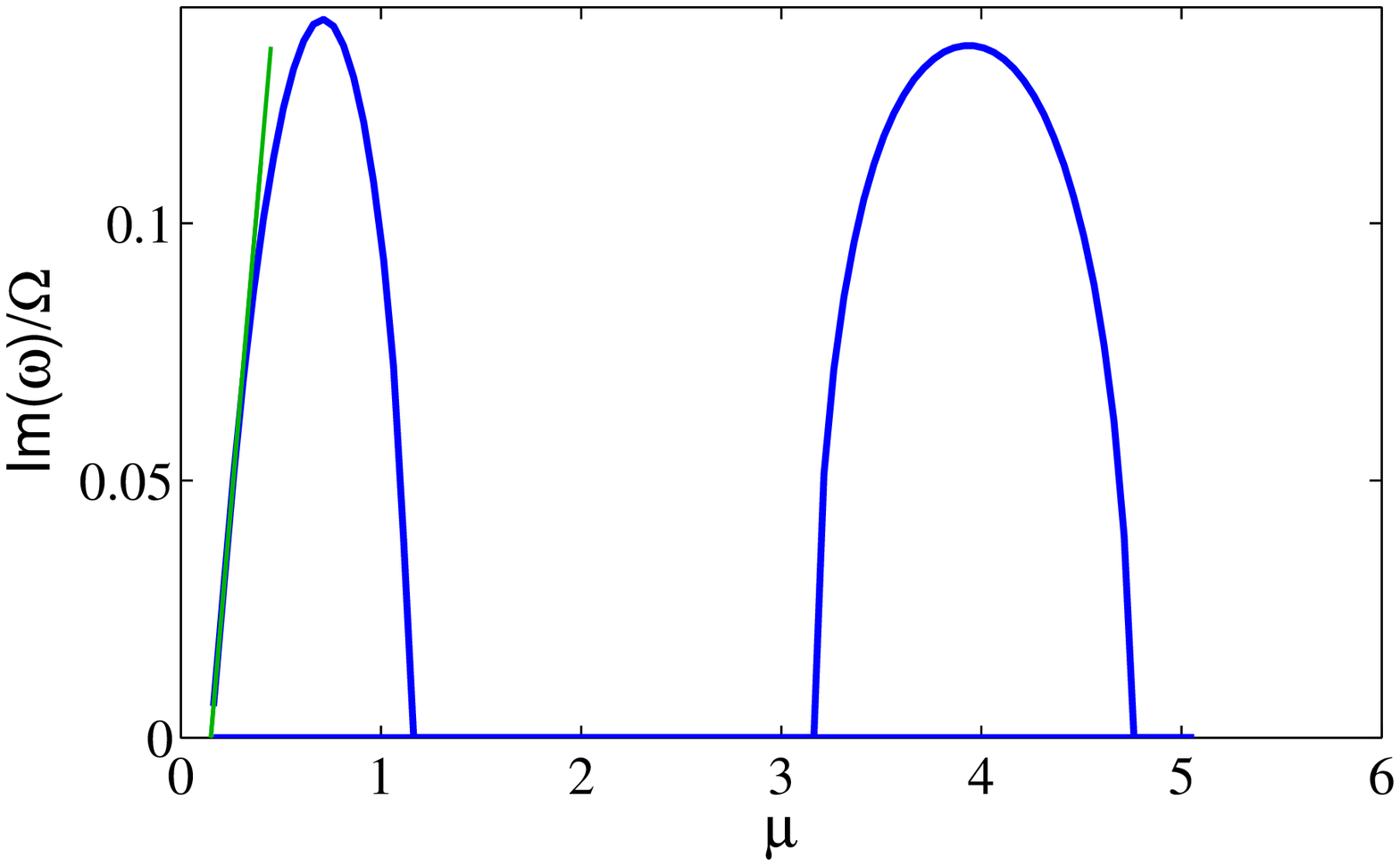} 
\includegraphics[width=8.00cm]{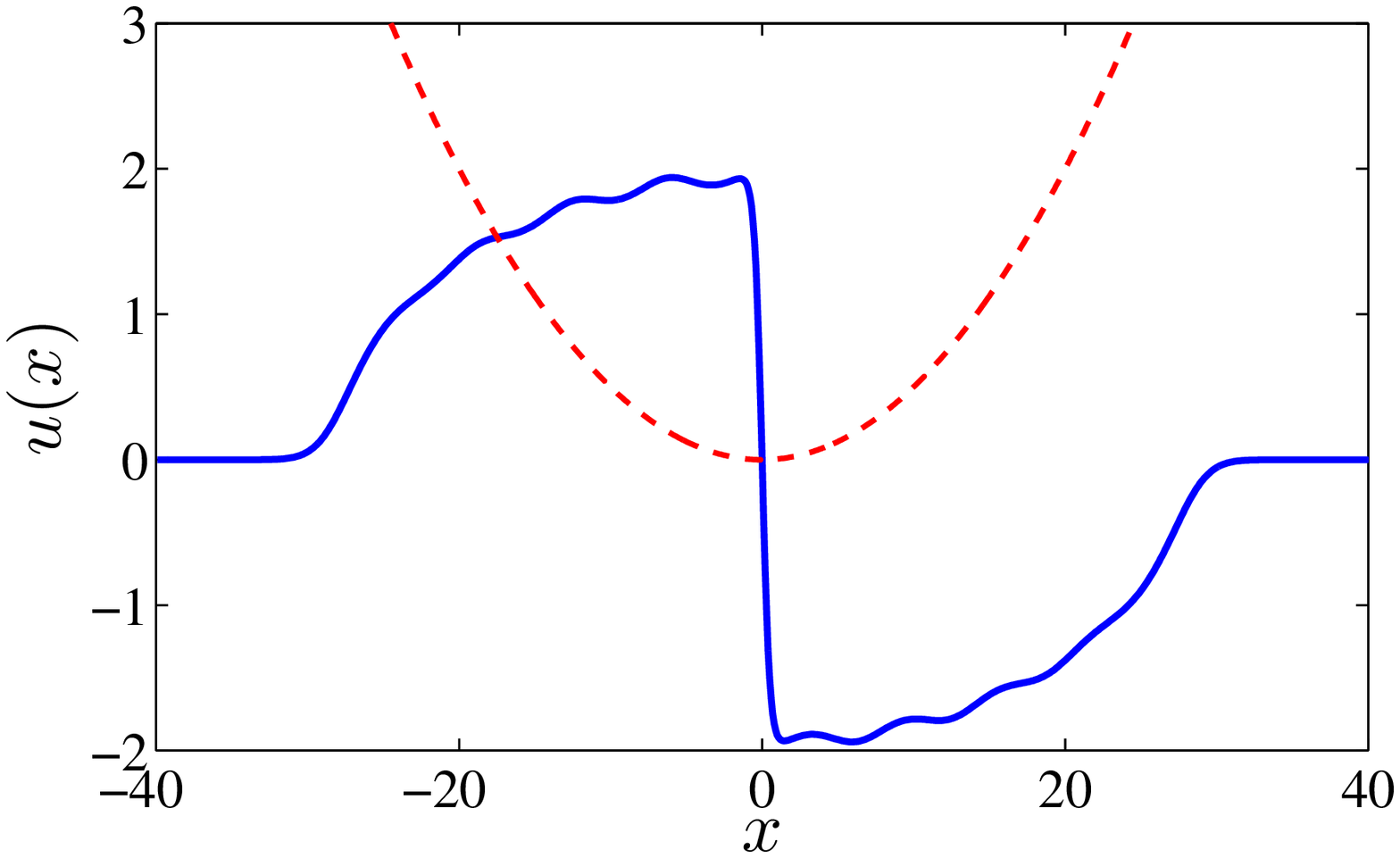}
\includegraphics[width=8.00cm]{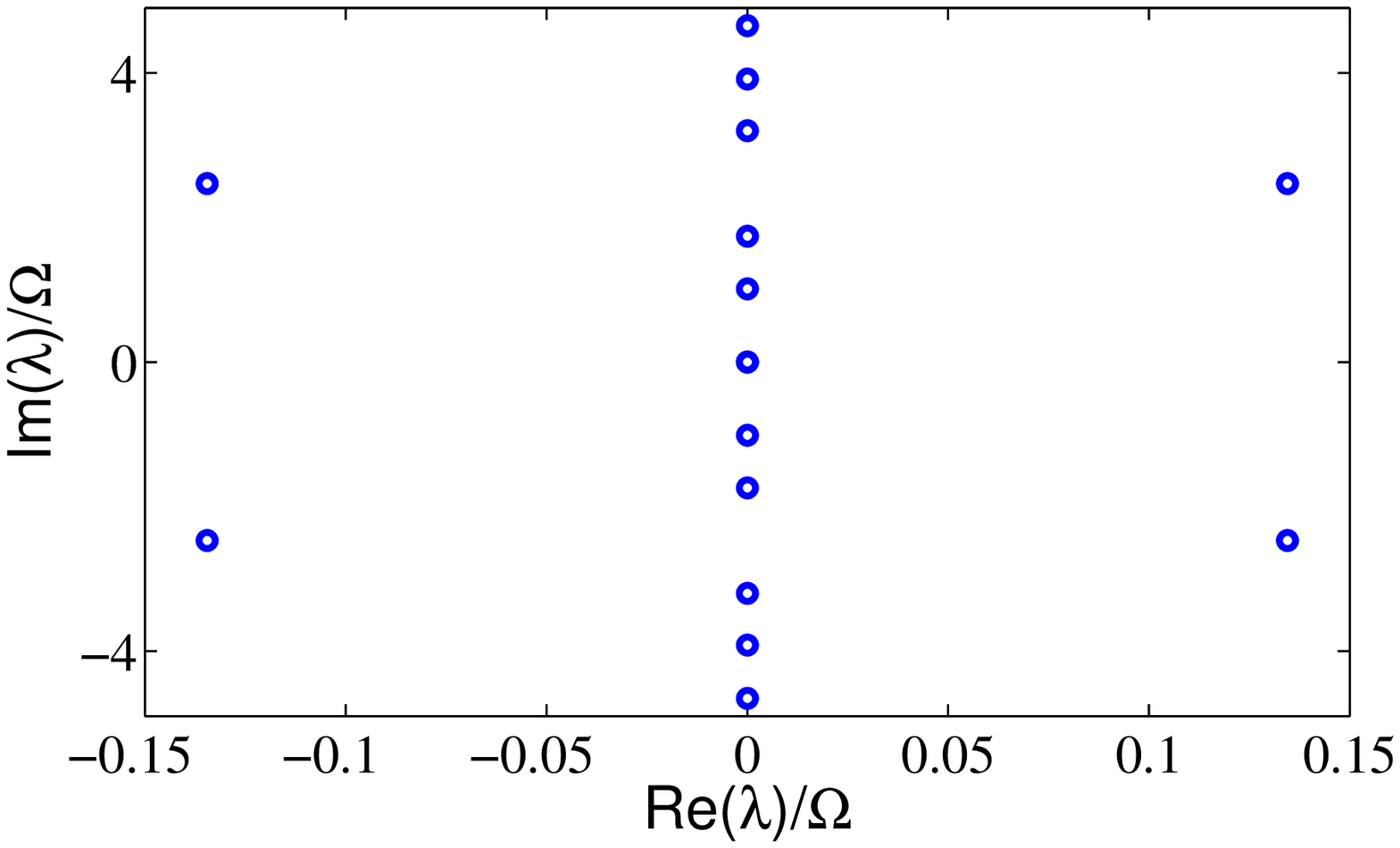} 
\caption{(Color online) The top left panel is similar to the previous ones, and illustrates
the real part of the eigenfrequency spectrum. 
The top right panel shows the imaginary part of the eigenfrequency spectrum. The bottom left
and right panels show, respectively, the solution (blue solid line), the
external parabolic trap (red (dashed) line) and the BdG spectrum (blue
circles in the bottom right), all given for $\mu=3.962$.}
\label{LCb}
\end{center}
\end{figure*}

Finally, in Fig.~\ref{f:spectrum}, we consider the case example of a three dark soliton state.
Now all three of the first frequencies
at $\Omega$, $2 \Omega$ and $3 \Omega$ are resonant. The first
pair however, associated with $n=4$ (positive energy) and $n=2$
(negative energy) for the case of $m=3$ does not lead to instability,
as again one of these modes is the dipolar one ($n=4$) and remains
invariant. The anomalous mode, on the other hand, moves according to:
\begin{eqnarray}
\omega= \Omega - \frac{19}{156} \left(\mu - \frac{7}{2} \Omega \right)
\label{pred7}
\end{eqnarray}
and decreases in frequency approximating in the TF limit, once again,
$\Omega/\sqrt{2}$ being associated with the in-phase motion of all
three dark solitons.
The other two pairs indeed do lead to instabilities 
near the linear limit. Indeed, the modes with $n=5$ (positive
energy) and $n=1$ (negative energy) lead to a resonance,
resolved by degenerate perturbation theory, according to the result:
\begin{eqnarray}
\omega= 2 \Omega + \left(\mu - \frac{7}{2} \Omega \right)
\frac{-163 \pm 7 \sqrt{71} i }{2352}.
\label{pred8}
\end{eqnarray}
Finally, the modes with $n=6$ (positive energy) and $n=0$
(negative energy) will be
degenerate in this case of $m=3$ at $3 \Omega$, and can
be captured via the degenerate perturbation methodology as:
\begin{eqnarray}
\omega= 3 \Omega + \left(\mu - \frac{7}{2} \Omega \right)
\frac{-769 \pm  \sqrt{124031} i }{16384}.
\label{pred9}
\end{eqnarray}
We note that for higher chemical potential, these complex
eigenfrequency quartets split into two pairs. In each of
these, the positive energy one follows the asymptotics
prescribed by Eq.~(\ref{tflimit}), while the negative
energy modes become associated with vibrations of the
dark solitons. The second anomalous mode relates to the out-of-phase
vibration of the outer dark solitons, while the middle
one remains quiescent (cf.~also a relevant experimental result 
in Ref.~\cite{ourmarkus3}). The third anomalous mode corresponds
to a more complex motion where the two outer solitons are
in-phase, while the middle one is out-of-phase with respect to them.
These are both captured accurately by Eq.~(31) of Ref.~\cite{coles}.

Figure~\ref{f:spectrum} confirms that the above theoretical predictions
are again in good agreement with numerical observations.
The decreasing tendency of the real part of the first three
eigenfrequencies (or eigenfrequency pairs) is accurately captured
by Eqs.~(\ref{pred7})--(\ref{pred9}). Equally importantly, the growth
rates relating to the two instabilities due to the resonances
at $2 \Omega$ and $3 \Omega$ that are also accurately represented near
the linear limit.

\begin{figure*}[tb] 
\begin{center}
\includegraphics[width=8cm]{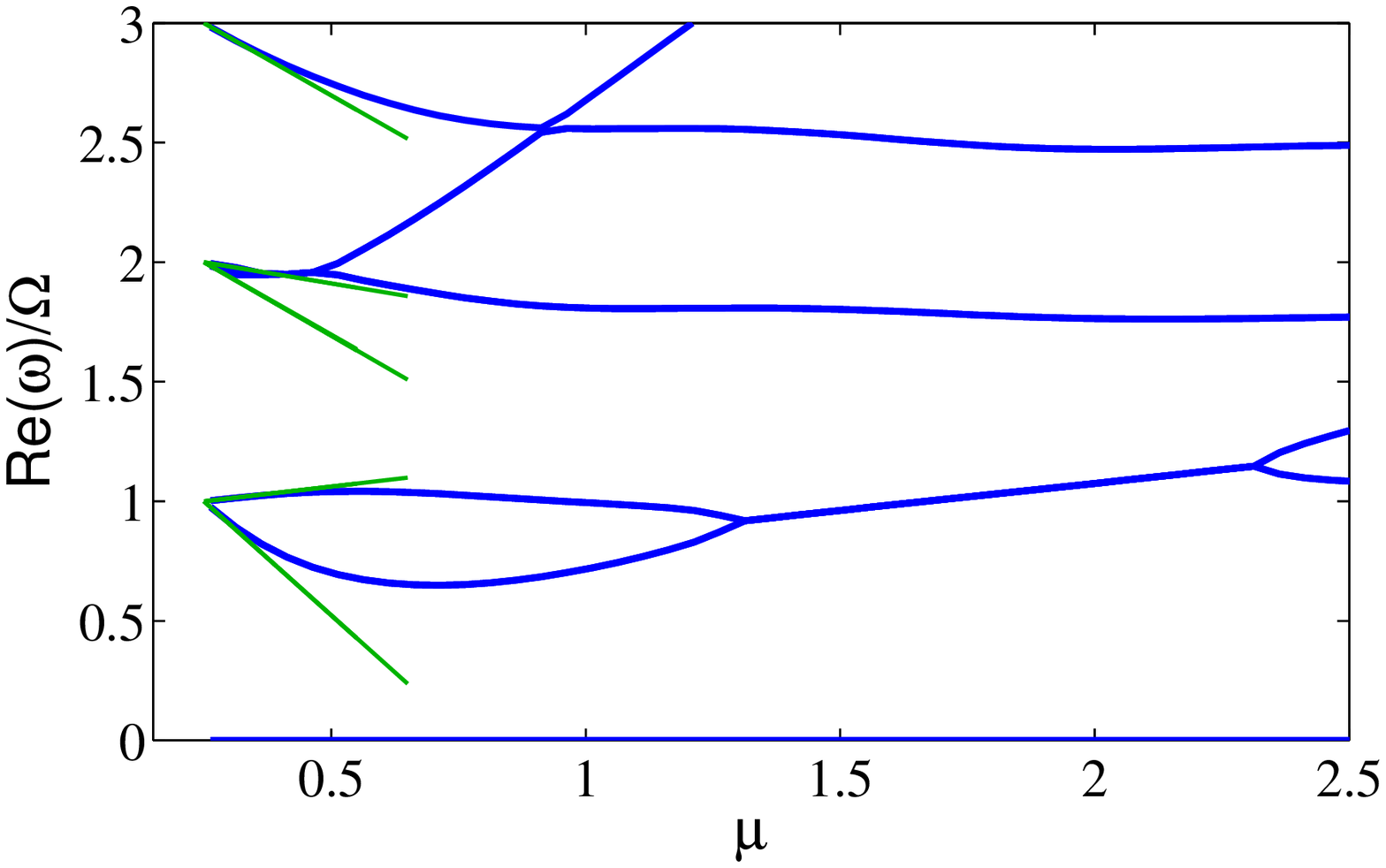}
\includegraphics[width=8cm]{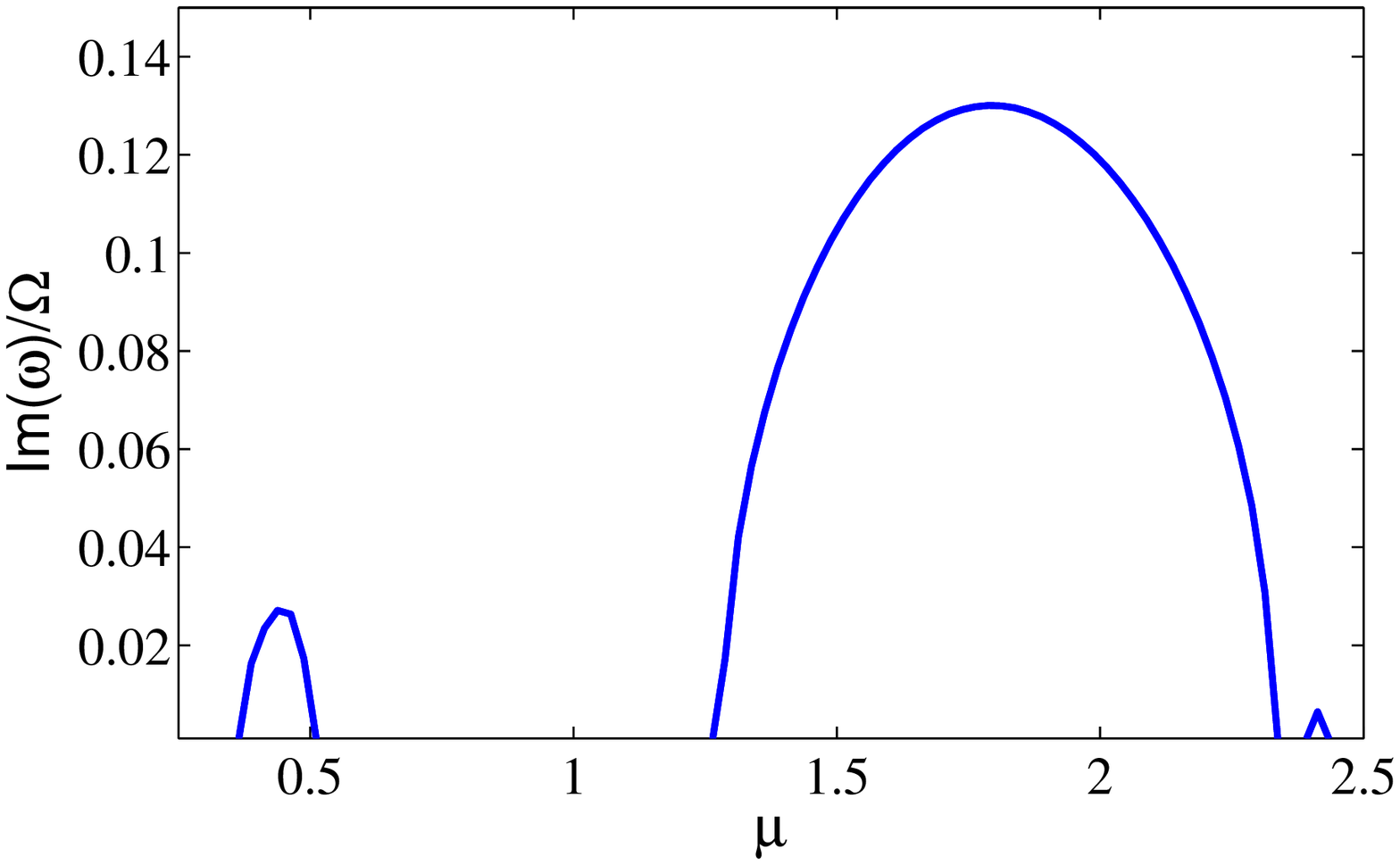} 
\includegraphics[width=8cm]{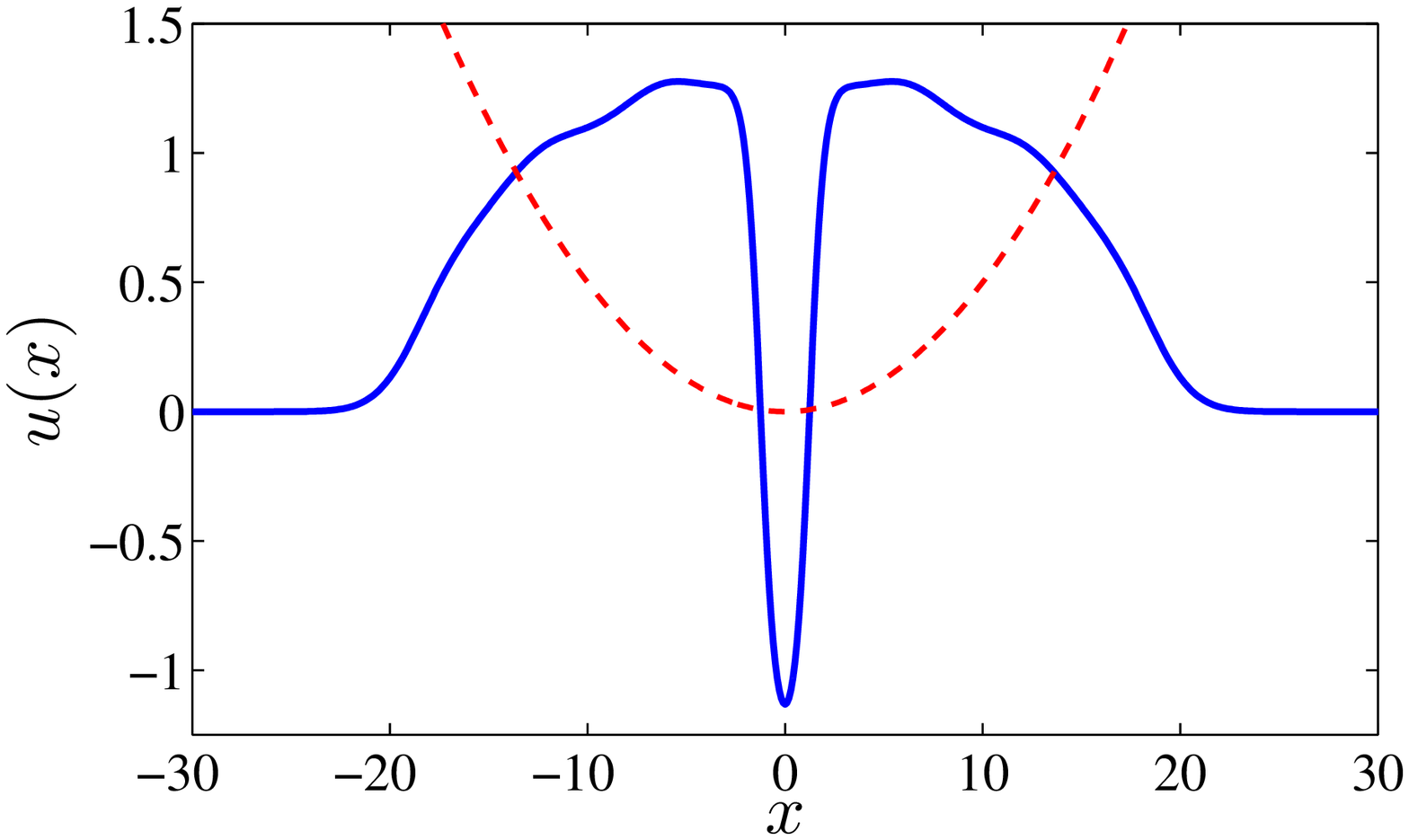}
\includegraphics[width=8cm]{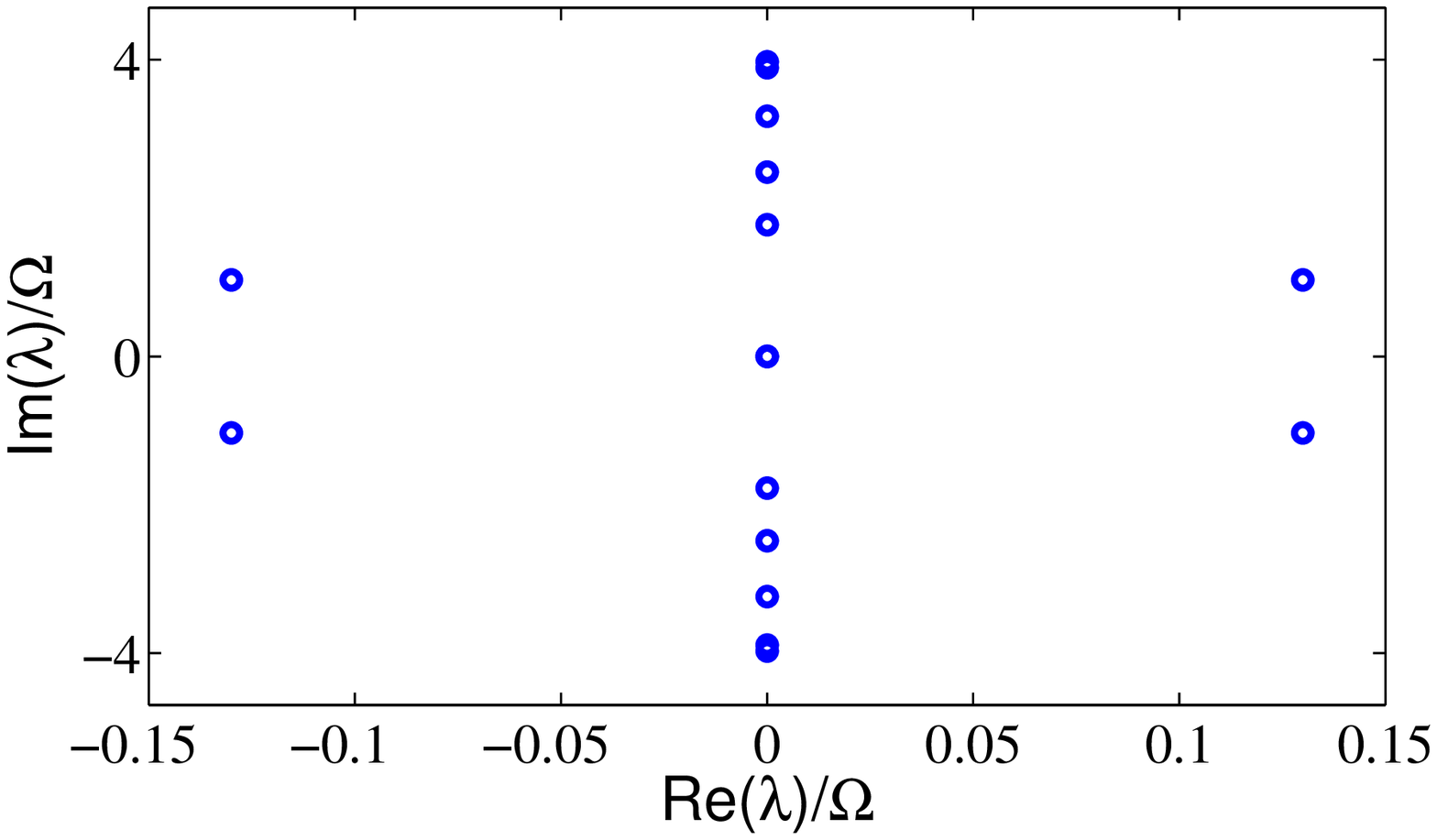} 
\caption{(Color online) Similar 
to Fig.~\ref{LCb}, but now for the
two soliton solution. The top panels show the real and imaginary
parts of the frequencies respectively, in good agreement
close to the linear limit to the theoretical predictions 
[dashed (red) line] of Eq.~(\ref{ci7}), while the bottom ones illustrate
the solution and its BdG spectrum for the case of $\mu=1.813$.
}
\label{LC2}
\end{center}
\end{figure*}

\begin{figure*}[tb] 
\begin{center}
\includegraphics[width=8cm]{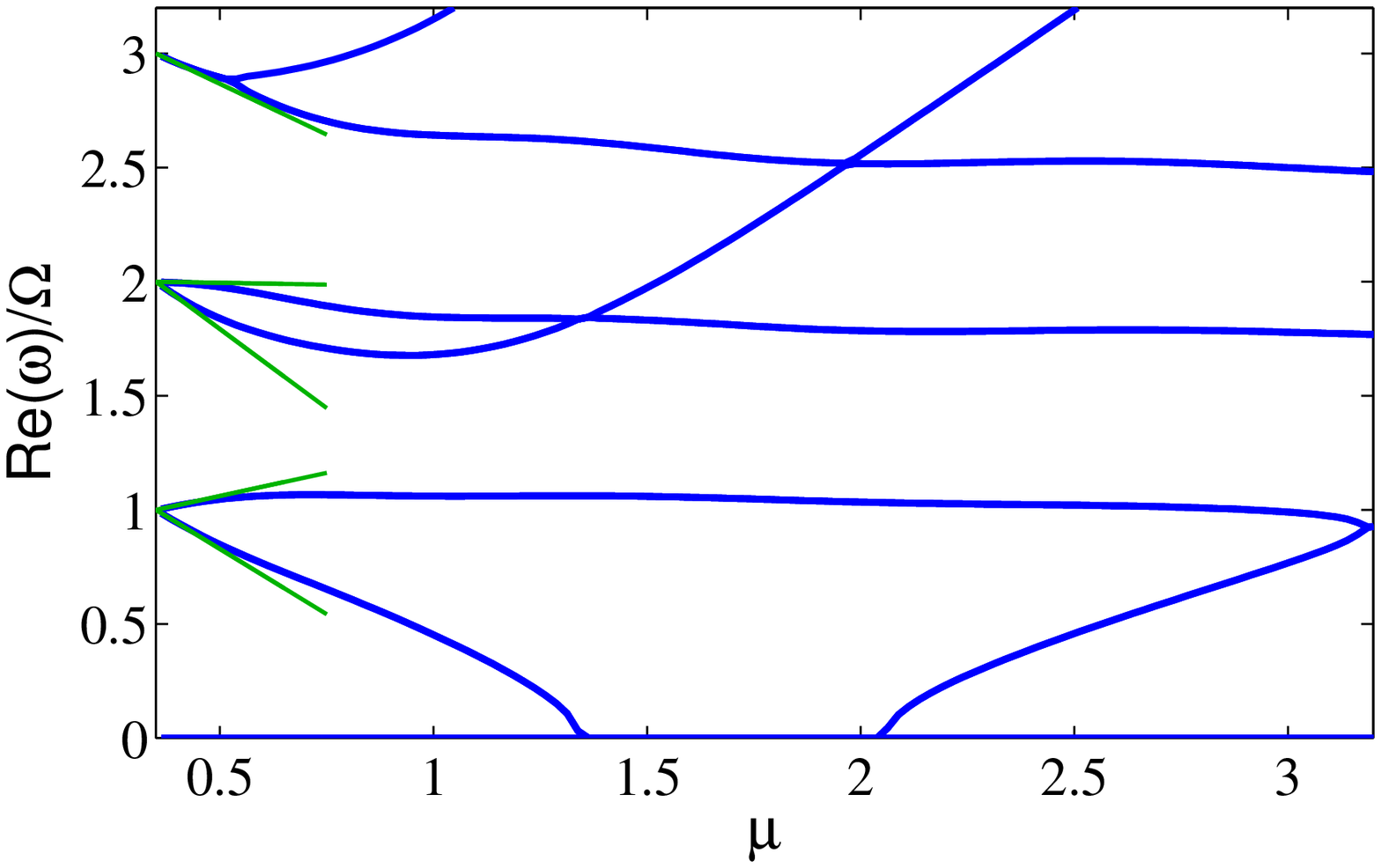}
\includegraphics[width=8cm]{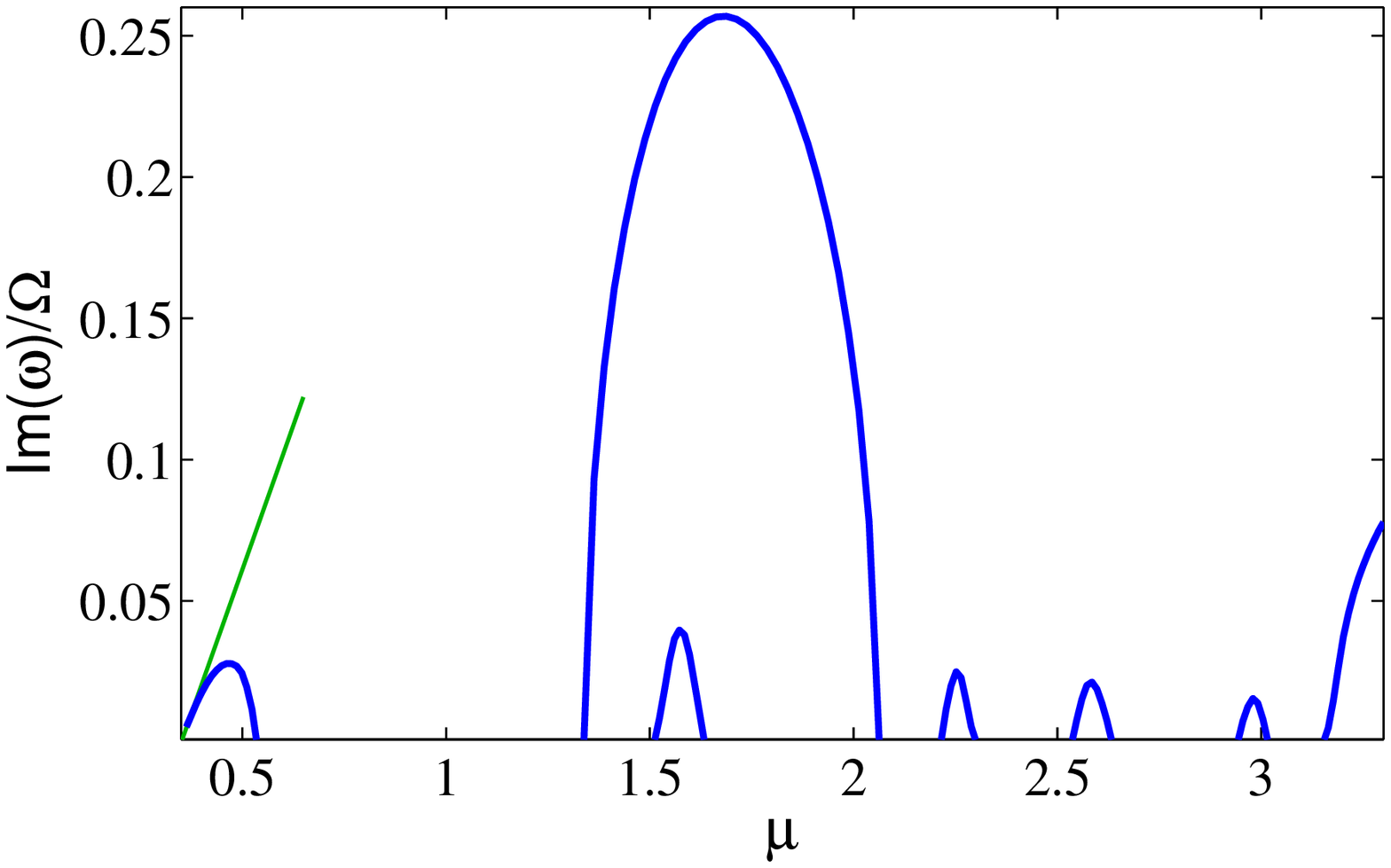} 
\includegraphics[width=8cm]{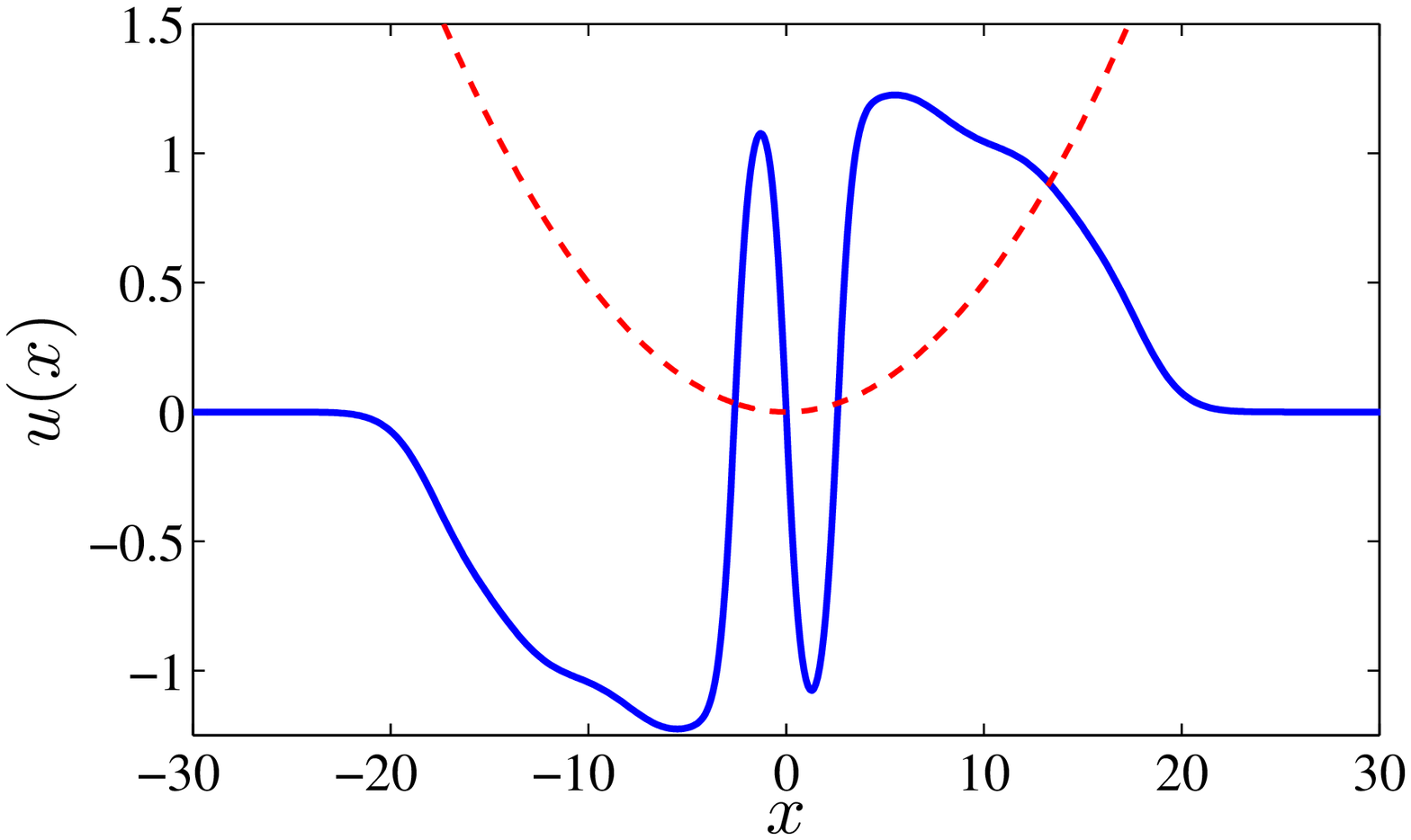}
\includegraphics[width=8cm]{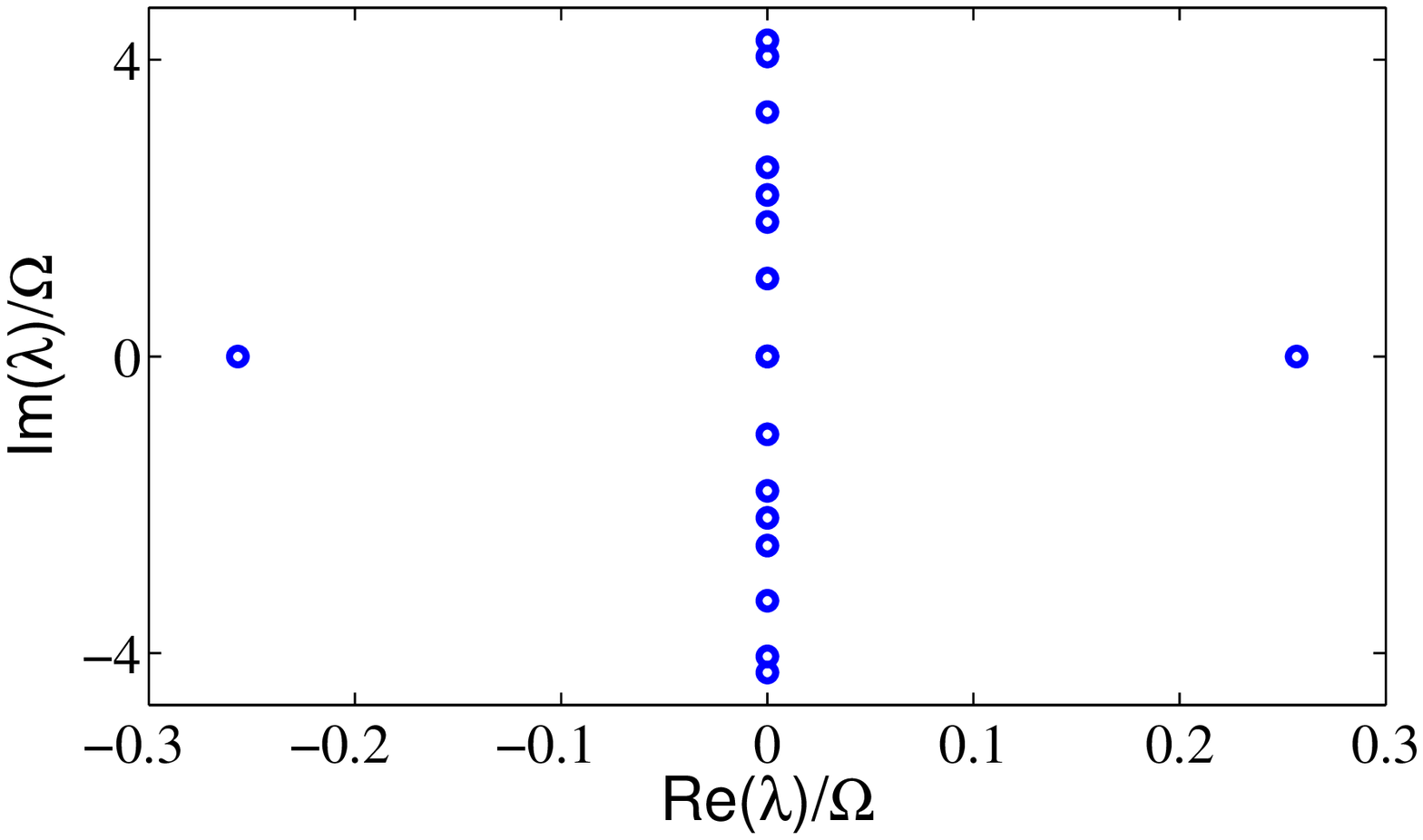} 
\caption{(Color online) Similar to the previous figures, but now
for the case of the three-soliton solution. Once again, the top
panels show the real and imaginary parts of the relevant eigenfrequencies,
now dominated by an exponential instability for $1.36 < \mu < 2.04$. The
theoretical predictions of Eq.~(\ref{ci8}) [dashed (red) lines]
are once again seen to
be accurate in the vicinity of the linear limit. The bottom panels
show the solution profile (bottom left) and spectral plane
$(\lambda_r, \lambda_i)$ for the case of $\mu=1.866$.
}
\label{LC3}
\end{center}
\end{figure*}

\subsection{Collisionally Inhomogeneous Case}

We now extend our considerations to the collisionally inhomogeneous
case, to examine the variations on the spectral picture that are
induced by the presence of $g(x)$. As indicated above, we will
primarily focus on a periodic variation, cf. Eq.~(\ref{g(x)}), 
as a principal building block towards more complex
such variations. Interestingly, the analytical considerations
of the previous section can still be carried out in the present
setting. However, the explicit formulae resulting are far more
tedious. For instance, in the case of the single dark soliton
the chemical potential correction $\mu_1$ is given by:
\begin{eqnarray}
\mu_1 = \frac{3 (2 + g_0) \Omega^2 - e^{-\frac{k^2}{2 \Omega}}
g_0 (k^4 + 6 k^2 \Omega + \Omega^2)}{8 \sqrt{2 \pi} \Omega^{3/2}}.
\label{ci1}
\end{eqnarray}
It is even more interesting to examine in this case the
corrections to the spectral frequencies. Recall that, as
before, at $\omega=\Omega$ two frequencies collide, the
positive energy one, with $n=2$, and the negative energy
one, with $n=0$, for this case of $m=1$. However, importantly,
the presence of the periodic variation {\it destroys} the
invariance associated with the dipolar mode. As a result,
the resonance may arise in this setting and is associated with
an eigenvalue given by:
\begin{eqnarray}
\omega= \Omega + \epsilon \omega_1^{{\rm cor}}
\label{ci2}
\end{eqnarray}
where $\epsilon= (\mu - 3 \Omega/2)/ \mu_1$ and $\omega_1^{{\rm cor}}$
is given by the following lengthy expression:
\begin{eqnarray}
 \omega_1^{{\rm cor}} = \frac{\Omega^{1/2}}{32 \sqrt{2 \pi}}
\left( A \pm  \sqrt{B} \right),
\label{ci3}
\end{eqnarray}
where
\begin{eqnarray}
A &\!\!=\!\!& -2 - g_0 + e^{-\frac{k^2}{2 \Omega}} g_0 (r^3 - 11 r^2 + 17 r + 1)
\nonumber
\\
B &\!\!=\!\!& (2+g_0)^2 +2 e^{-\frac{k^2}{2 \Omega}} g_0 (2+g_0)
(3 r^3 - 13 r^2 - 5 r - 1),
\nonumber
\\
&\!\!+\!\!& e^{-\frac{k^2}{ \Omega}} g_0^2
(r^2-2 r - 1)
(r^4 - 12 r^3 +36 r^2 - 8 r - 1),
\nonumber
\\
r&\!\!=\!\!&k^2/\Omega.
\nonumber
\end{eqnarray}
Furthermore, it is straightforward to observe that in the homogeneous
limit of $g_0 \rightarrow 0$, this result retrieves Eq.~(\ref{pred1}), 
as well as the dipolar mode with $\omega=\Omega$.
It is relevant to also point 
out that the above expression, due to the presence of the radical, 
provides implicitly conditions (through the zero crossing of $B$) 
under which the two real pairs colliding at this frequency
lead to a complex eigenfrequency quartet. However, it is also
evident that while these expressions are available in explicit
form, they are particularly tedious and, hence, we will only
provide them for the single soliton case. The other two modes
associated with $n=3$ and $n=4$, at $2 \Omega$ and $3 \Omega$
respectively, are non-resonant and yield eigenfrequency corrections,
as follows:
\begin{eqnarray}
\omega_1^{{\rm cor}} &\!\!=\!\!& \frac{\Omega^{1/2}}{96 \sqrt{2 \pi}}
\left[-3 (2 +g_0) \phantom{e^{-\frac{k^2}{2 \Omega}} } \right.
\nonumber
\\
\nonumber
&\!\!+\!\!&
\left.  e^{-\frac{k^2}{2 \Omega}} g_0 (-r^4 + 16 r^3
-54 r^2 + 24 r +3) \right], ~~ {\rm and}
\label{ci4}
\\
\omega_1^{{\rm cor}} &\!\!=\!\!& \frac{\Omega^{1/2}}{768 \sqrt{2 \pi}}
\left[-63 (2 +g_0)  \phantom{e^{-\frac{k^2}{2 \Omega}} } \right.
\nonumber
\\
\nonumber
&\!\!+\!\!&
\left.  e^{-\frac{k^2}{2 \Omega}} g_0 (r^5 - 21 r^4 +
126 r^3 -246 r^2 + 45 r +63) \right].
\label{ci5}
\end{eqnarray}

Let us now compare these predictions with the numerical results, as shown in Fig.~\ref{LCb}.
The bottom left panel of the figure shows an example of the single
soliton profile, at a relatively large value of $\mu=3.962$, in order
to clearly illustrate the effect of the periodic variation towards
the solitonic structure inside the parabolic trap (also shown).
The rest of the parameters are chosen as $\Omega=0.1$, $g_0=0.1$
and $k=0.5$. These three parameters will indeed be fixed hereafter
to those values, although it is clear from Eqs.~(\ref{ci1})--(\ref{ci5})
that our methodology can tackle general parametric combinations.
In this case, it turns out that the quantity $B$ in Eq.~(\ref{ci3})
is negative, hence the relevant pair of modes predicted in
Eq.~(\ref{ci3}) is complex for values of $\mu$ near the linear
limit. This, as well as the overall
trend of the eigenfrequency variations in Eq.~(\ref{ci3})
is accurately predicted by the theory (cf.~top left panel of
Fig.~\ref{LCb}). Moreover, the imaginary 
part of the relevant eigenfrequency (i.e., the growth rate of
the associated instability) is also accurately captured, as shown in the
top right panel of the figure. 

However, in the same panel, it can be
seen that the existence of a resonance and an instability of
even the single dark soliton (entirely contrary to what is the
case in the collisionally homogeneous environment) is {\it not} the
only feature of the spectrum. In addition to that instability
(occurring, in this case, for $ 0.15 < \mu < 1.16$), there is another instability
that arises for higher values of the chemical potential i.e.,
for $3.16 < \mu < 4.76$. This instability 
can be straightforwardly inferred from the bottom right panel, 
showing the spectral plane $(\lambda_r, \lambda_i)$ of the eigenvalues
$\lambda= \lambda_r + i \lambda_i$. There, it can be seen that the anomalous
mode as it increases collides with the third pair of eigenfrequencies
in the vicinity of $ 3 \Omega$ producing this instability.
It is important to highlight once again that, for $g_0=0$, the single
dark soliton is generically spectrally stable, hence all these
instability features do not arise.

In the case of the two soliton state,  
for the above mentioned
choice of the parameters, the stability and existence results
are shown in Fig.~\ref{LC2}. Here, the correction to the chemical
potential is given by:
\begin{eqnarray}
\mu_1&=&\frac{\Omega^{1/2}}{128 \sqrt{2 \pi}}
\left[41 (2 + g_0)\phantom{e^{-\frac{k^2}{2 \Omega}} } \right.
\nonumber
\\
\nonumber
&&
\left.- e^{-\frac{k^2}{2 \Omega}} g_0 (r^4 - 20 r^3 +
114 r^2 -172 r + 41) \right],
\label{ci6}
\end{eqnarray}
yet the rest of the analytical expressions is, arguably, too
complex to be presented 
here. Instead, for reasons of completeness,
we 
provide the theoretical predictions for the numerically selected
values of $\Omega=0.1$, $g_0=0.1$
and $k=0.5$. We find that:
\begin{eqnarray}
\omega&=& 0.1 + \epsilon (-0.0159, 0.0021),
\nonumber
\\
\omega&=& 0.2 + \epsilon (-0.0102, -0.0030),
\label{ci7}
\\
\nonumber
\omega&=&0.3 -0.0101 \epsilon,
\end{eqnarray}
with $\epsilon= (\mu -0.25)/ 0.0834$.
The first two pairs in Eq.~(\ref{ci7}) are meant to indicate
that the two eigenfrequency pairs for these parameter
values are not resonant {\it immediately after} the linear limit.
In fact, in one of the cases their sign of variation is opposite
(for those bifurcating at $\omega= \Omega$). The other pair
at $ 2 \Omega$ leads to an instability shortly after the linear
limit [although in agreement with Eq.~(\ref{ci7}) this does not
happen immediately]. It can be seen in the top left panel of
Fig.~\ref{LC2} that, once again, the near linear stability predictions
are fairly accurate here. Naturally, for larger values of $\mu$,
more complex phenomena may arise as, e.g., the collision of the
first positive and the first negative energy modes observed 
in the interval $1.312 < \mu < 2.313$.
This is also showcased in the bottom panel examples of the solution
profile and its BdG spectrum, shown for $\mu=1.813$.

However, it is important
to highlight again here the differences that the collisional
inhomogeneity may bring to the collisionally homogeneous picture.
Namely, in the collisionally homogeneous setting, the dark soliton pair would be
unstable in the immediate vicinity of the linear limit, while
here the presence of a spatial modulation of the nonlinearity may
(parametrically) ``delay'' this instability, while it may induce
other ones in different parametric ranges (similarly to what 
we 
found also in the single dark soliton case).

Finally, we also consider the continuation over the chemical
potential of the branch of three solitons in a collisionally
inhomogeneous setting. In this case, the results are presented
in Fig.~\ref{LC3}. The corresponding theoretical predictions
near the linear limit are
\begin{eqnarray}
\omega&=& 0.1 + \epsilon (-0.0085, 0.0030),
\nonumber
\\
\label{ci8}
\omega&=& 0.2 + \epsilon (-0.0103, -0.0002),
\\
\nonumber
\omega&=&0.3 +  \epsilon (-0.0066 \pm 0.0030 i),
\end{eqnarray}
with $\epsilon= (\mu -0.35)/ 0.0746$.
Despite the fairly complicated nature of the relevant
structure, once again  we can see in Fig.~\ref{LC3} that
the analytical predictions do well in capturing the
numerical results. This is true both for the first
two sets of modes, where despite the potential for resonance,
it is instead predicted that for the considered parameter
values, no such resonance exists. It is also true for the third
set of pairs that still leads to a resonance and to a complex
eigenvalue quartet, accurately captured both in the
real part (top left panel of the figure) and in the imaginary part
(top right panel showing the instability growth rate)
of the corresponding eigenfrequency.

However, in this case too, there are features that are
arguably somewhat unexpected and, in any case, fundamentally
distinct from the collisionally homogeneous case.
In particular, we observe that the lowest anomalous eigenfrequency
decreases in value until it collides with the origin, resulting
in the relevant eigenfrequency pair exiting as {\it purely}
imaginary for the parametric interval $1.36 < \mu < 2.04$.
This type of instability was {\it never} observed
for the configurations considered in the collisionally homogeneous
case. The instability is also illustrated in the bottom panels of the figure
representing the solution profile and the BdG stability analysis
for $\mu=1.688$.

\begin{figure}[tb] 
\begin{center}
\includegraphics[width=8cm]{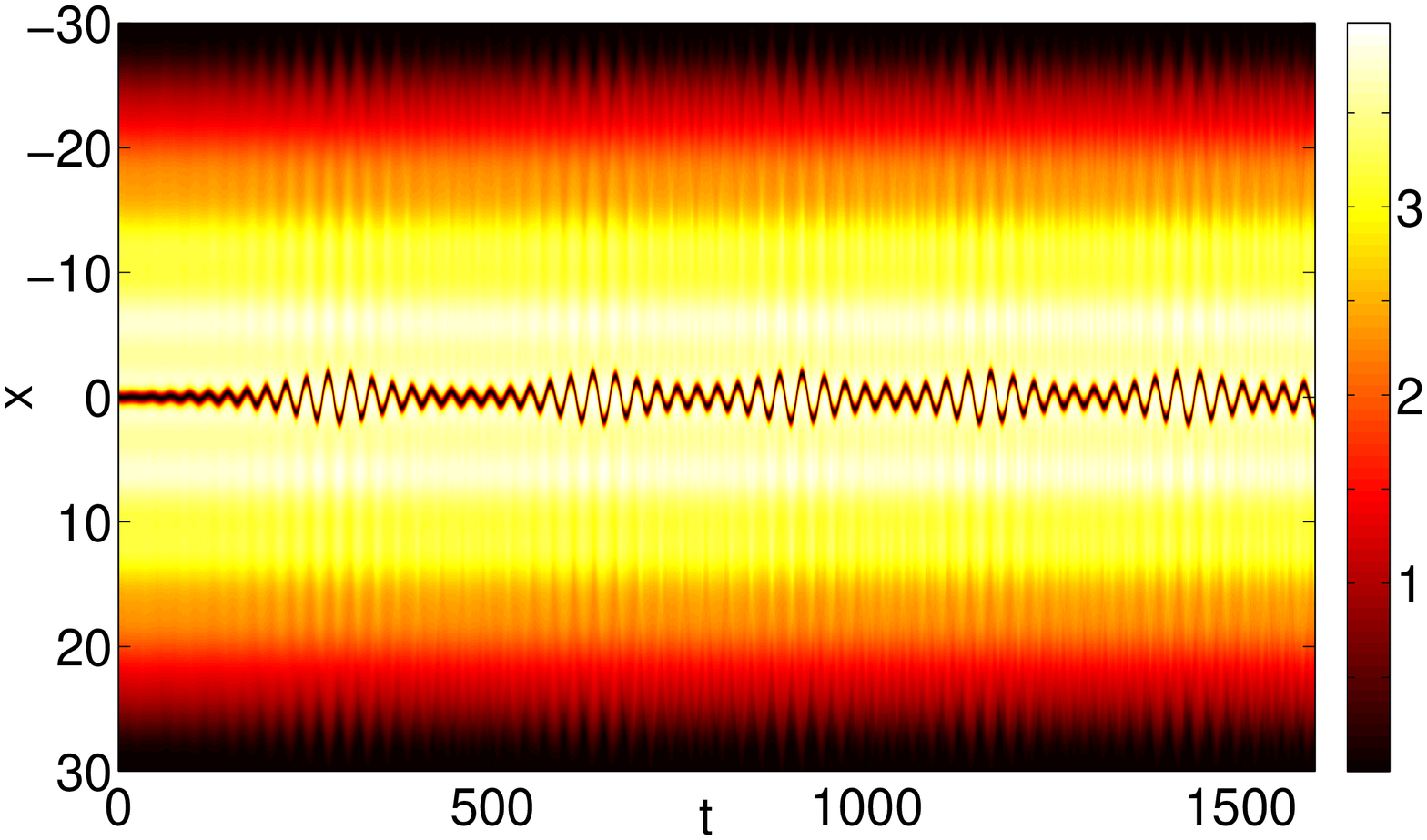}\\[1.0ex]
\includegraphics[width=8cm]{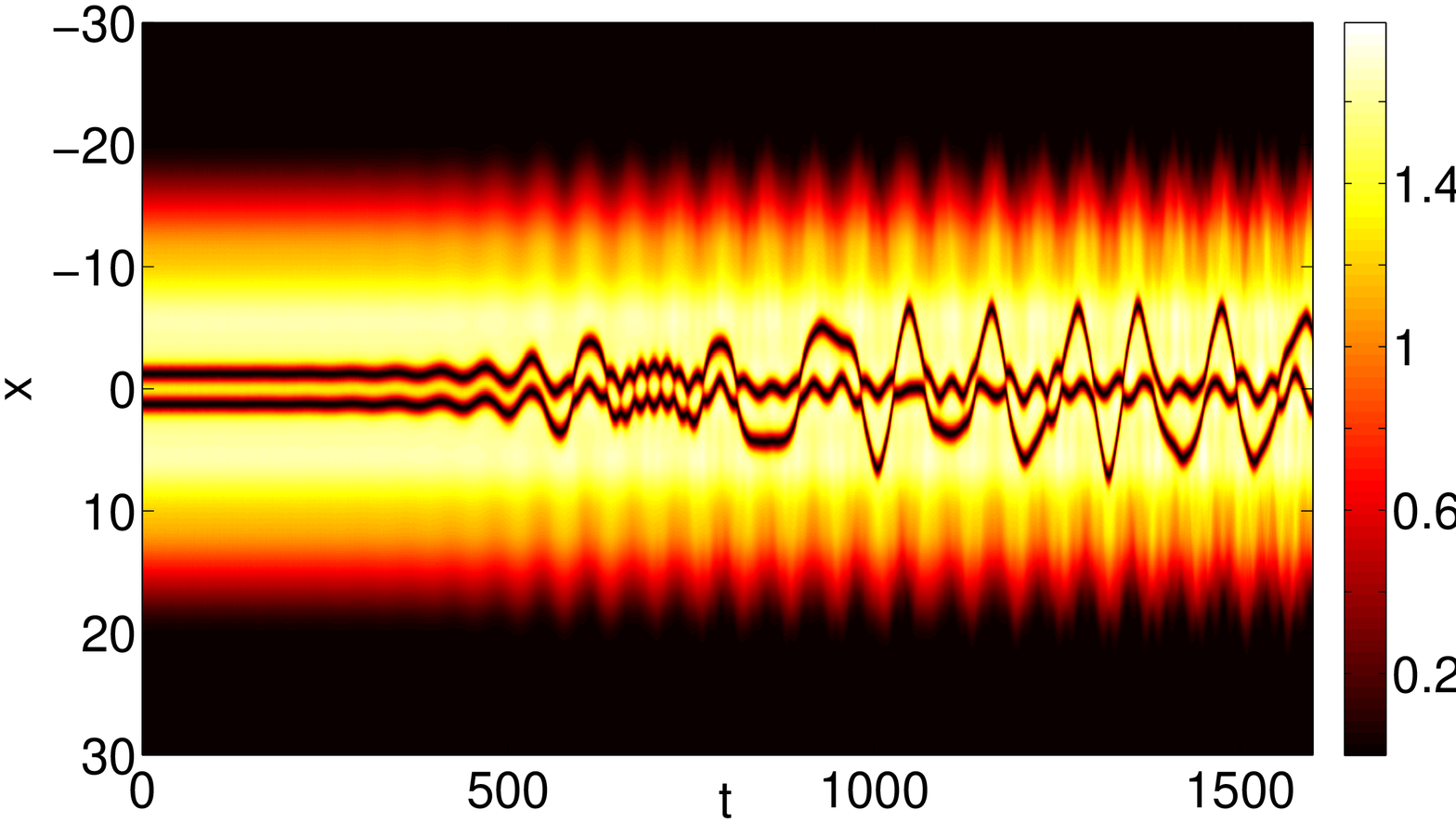}\\[1.0ex]
\includegraphics[width=8cm]{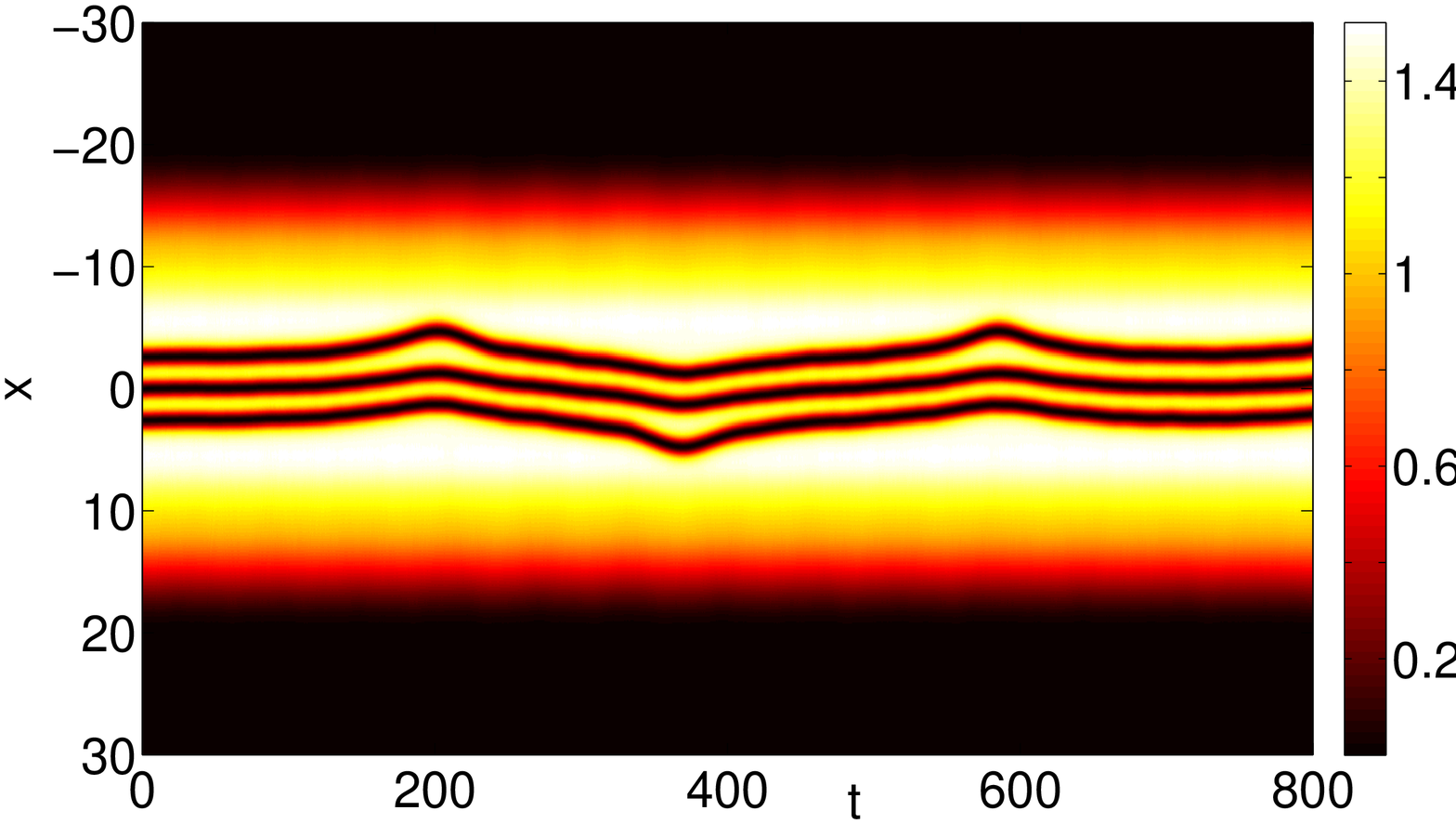}
\caption{(Color online) Space-time density contour plots illustrating
the unstable evolution of the case examples
shown in the bottom panels of Figs.~\ref{LCb},
\ref{LC2} and~\ref{LC3}, respectively for
one dark soliton and $\mu=3.962$ (top), two solitons
and $\mu=1.813$ (middle), as well as three solitons
and $\mu=1.688$ (bottom).
}
\label{LC4}
\end{center}
\end{figure}

The above results clearly show that the collisionally inhomogeneous
case presents a number of features (illustrated in Figs.~\ref{LCb}-\ref{LC3})
that are absent in the context of the collisionally homogeneous
setting. Some of the prototypical ones of these features are
worth dynamically exploring, as is depicted in Fig.~\ref{LC4}.
Our earlier work in the collisionally homogeneous setting~\cite{ourmarkus3}
demonstrated that the oscillatory instabilities of the two- and three-dark solitons 
there corresponded to modes of energy exchange between the 
solitons and the background 
-- cf.~Figs~\ref{LC} and~\ref{f:spectrum}. 
In the top panel
of Fig.~\ref{LC4}, we encounter an oscillatory instability even
for a single dark soliton in line with the results of Fig.~\ref{LCb}
(an instability absent for homogeneous settings). This instability is seen
to lead to an oscillation of the solitary wave and an energy exchange
with a corresponding mode of the background (as is implied by the associated
resonance). Nevertheless, the dynamics of the solitary wave remains
oscillatory for a long time, without any other major features arising
in the dynamics for the soliton shown in Fig.~\ref{LCb}, for
$\mu=3.962$. 

In the middle panel of Fig.~\ref{LC4}, we see an instability
of the two soliton state arising from the collision of the anomalous
mode with $n=1$ with the ``former dipolar'' mode of $n=3$. Note that
this instability for $\mu=1.813$ (the soliton is shown in the bottom
panel of Fig.~\ref{LC2}) would be {\it absent} in the collisionally
homogeneous limit. Namely, there is {\it no} instability in the latter
limit for the soliton in-phase motion. Yet, the presence of collisional
inhomogeneity destroys the invariance associated with the dipolar motion
and allows the resonant collision of the in-phase mode with the dipolar
one and the corresponding emergence of an oscillatory growth of
the in-phase motion, as depicted in the middle panel of Fig.~\ref{LC4}.
Eventually, this mode undergoes a dramatic modification in its character,
as a result of the instability resulting in one of the dark solitons
executing smaller amplitude oscillations near the center, while the
other one executes large amplitude oscillations, reaching the 
rims of the condensate. 
This behavior is reminiscent, albeit involving a smaller number of solitons,  
of the quantum's Newton cradle behavior observed in larger dark soliton chains, 
as described in Ref.~\cite{NewtonCradle}.
It is interesting to note that some of dark soliton collisions in our setting
render evident the repulsive nature of the 
interaction, while others illustrate the finite
height of the repulsive barrier, given that the solitons effectively
can go through each other (if they possess sufficient kinetic energy,
as, e.g., in the collisions around $t=1340$ or $t=1500$).

Lastly, yet another example of an instability absent in the
collisionally homogeneous limit is showcased for $\mu=1.688$
in the bottom panel of Fig.~\ref{LC4}. This is the genuinely
exponential instability arising in the case of the three-soliton
solution, shown in the bottom panel of Fig.~\ref{LC3}.
Here, we can see that the mode is no longer resonant with another
mode, but merely reflects the exponential growth eventually
associated with the in-phase motion of the three solitons,
as expected based on the spectrum of Fig.~\ref{LC3}.
As an aside, it is interesting to point out here the more
curved trajectory of the outer solitons as they reach the outer
rims of the condensate, presumably because they feel more
intensely there the periodic variation of the nonlinearity coefficient.

\section{Conclusions and Future Work
\label{sec:conclu}}

In the present work we explored the stability -- and partly the associated
dynamics -- of single and multiple matter-wave dark solitons in BECs confined in parabolic traps. 
We focused on the contradistinction between the collisionally homogeneous
case, of a constant nonlinearity coefficient, and the collisionally
inhomogeneous one, of a spatially-varying nonlinearity
coefficient. As our approach of choice, we utilized the vicinity
of the linear limit 
(corresponding to the quantum harmonic oscillator problem) enabling
us to employ a variant of the degenerate perturbation theory for
this Hamiltonian system. 
Our analytical approach provided 
a systematic handle on the nature of the (positive energy, or negative energy --anomalous--) 
modes. This also allowed us to quantitatively characterize these 
modes, at least in the vicinity of the linear limit. In all the
cases considered, this was found to be a useful tool for understanding
both the motion of the relevant modes, as well as for assessing
their potential for yielding oscillatory instabilities associated
with complex bifurcations near this limit. 
Supplemented with an understanding
(at least in the collisionally homogeneous case) of the spectral
picture for large chemical potentials,   
our analytical approximations completed the full range of parametric regimes that
are analytically tractable. On the other hand, 
the numerical computations provided
a smooth interpolation between these asymptotic cases.

Our investigation revealed a significant wealth of differences
between the collisionally homogeneous and the collisionally
inhomogeneous cases. For instance, in the latter, even the single
dark soliton structure might become unstable, while it is always
stable in one-dimensional collisionally homogeneous setting. 
Also, two-soliton
states may feature an in-phase resonant instability, which is
again excluded by symmetry in the homogeneous case. Finally,
for three solitons even an exponential instability of their
in-phase mode is possible, again a trait that is not encountered
in the homogeneous case. For these instabilities, given their
previously unexplored nature, we examined the ensuing dynamics
through direct numerical simulations.

There are numerous questions that are of interest
to explore in future work. On the one hand, tackling the
large chemical potential limit of the inhomogeneous case, 
and understanding the relevant Bogolyubov-de Gennes spectrum
in that limit in as much generality as possible, would certainly
be of interest. On the other hand, in the case of higher-dimensional 
structures (such as 
vortex states) in a single component,
or in that of structures arising in multicomponent systems 
(even in one dimension, such as the dark-bright soliton), 
developing a perturbative count of the relevant eigenvalues and of their dependence 
on the chemical potential (or potentials, as in multiple components,
there are multiple such parameters) would be a direction 
also very worthwhile to pursue. Work along some of 
these directions is currently in progress and will be
reported in future publications.

\begin{acknowledgments}
  P.G.K.~gratefully acknowledges the support of NSF-DMS-1312856,
and NSF-PHY-1602994, the Alexander von Humboldt Foundation,
  as well as from
the 
the ERC under FP7, Marie
Curie Actions, People, International Research Staff Exchange Scheme (IRSES-605096).
R.C.G.~gratefully acknowledges the support of NSF-DMS-1309035.
\end{acknowledgments}

\end{document}